%% file: main.tex
\documentclass[sigconf,screen]{acmart}  %

\usepackage[utf8]{inputenc}
\usepackage{amsmath}
\usepackage{hyperref}
\usepackage{xcolor}
\usepackage{color, colortbl}
\usepackage{breakcites}
\usepackage{tabularx}
\usepackage{booktabs}

\newcommand{\system}{\textsc{Captivate!}}

\copyrightyear{2022}
\acmYear{2022}
\setcopyright{acmlicensed}\acmConference[CHI '22]{CHI Conference on Human Factors in Computing Systems}{April 29-May 5, 2022}{New Orleans, LA, USA}
\acmBooktitle{CHI Conference on Human Factors in Computing Systems (CHI '22), April 29-May 5, 2022, New Orleans, LA, USA}
\acmPrice{15.00}
\acmDOI{10.1145/3491102.3501865}
\acmISBN{978-1-4503-9157-3/22/04}

\begin{document}
\title{Captivate! Contextual Language Guidance for Parent--Child Interaction}

\author{Taeahn Kwon}
\affiliation{
  \institution{Columbia University}
  \city{New York}
  \state{NY}
  \country{USA}
}
\email{taeahn.kwon@columbia.edu}
\authornote{Work done at Seoul National University}

\author{Minkyung Jeong}
\affiliation{
  \institution{Seoul National University}
  \city{Seoul}
  \country{South Korea}
}
\email{jmk7895@snu.ac.kr}
\authornote{Human-Centered Computer Systems Lab, Dept. of Computer Science and Engineering}

\author{Eon-Suk Ko}
\affiliation{
  \institution{Chosun University}
  \city{Gwangju}
  \country{South Korea}
}
\email{eonsukko@chosun.ac.kr}
\authornote{Child Language Lab, Dept. of English Language and Literature}

\author{Youngki Lee}
\affiliation{
  \institution{Seoul National University}
  \city{Seoul}
  \country{South Korea}
}
\email{youngkilee@snu.ac.kr}
\authornotemark[2]

\begin{abstract}
To acquire language, children need rich language input.
However, many parents find it difficult to provide
children with sufficient language input, which
risks delaying their language development.
To aid these parents, we design \system, the first system that
provides contextual language guidance to parents during play.
Our system tracks both visual and spoken language cues to infer
targets of joint attention, enabling the real-time suggestion
of situation-relevant phrases for the parent.
We design our system through a
user-centered process with immigrant families---a highly vulnerable
yet understudied population---as well as professional speech language
therapists. Next, we evaluate \system{} on parents with children
aged 1--3 to observe improvements in responsive language use. 
We share insights into developing contextual guidance technology
for linguistically diverse families\footnote{We share our code at \url{https://hcs.snu.ac.kr/captivate}.}.
\end{abstract}



\begin{teaserfigure}
  \includegraphics[width=\textwidth]{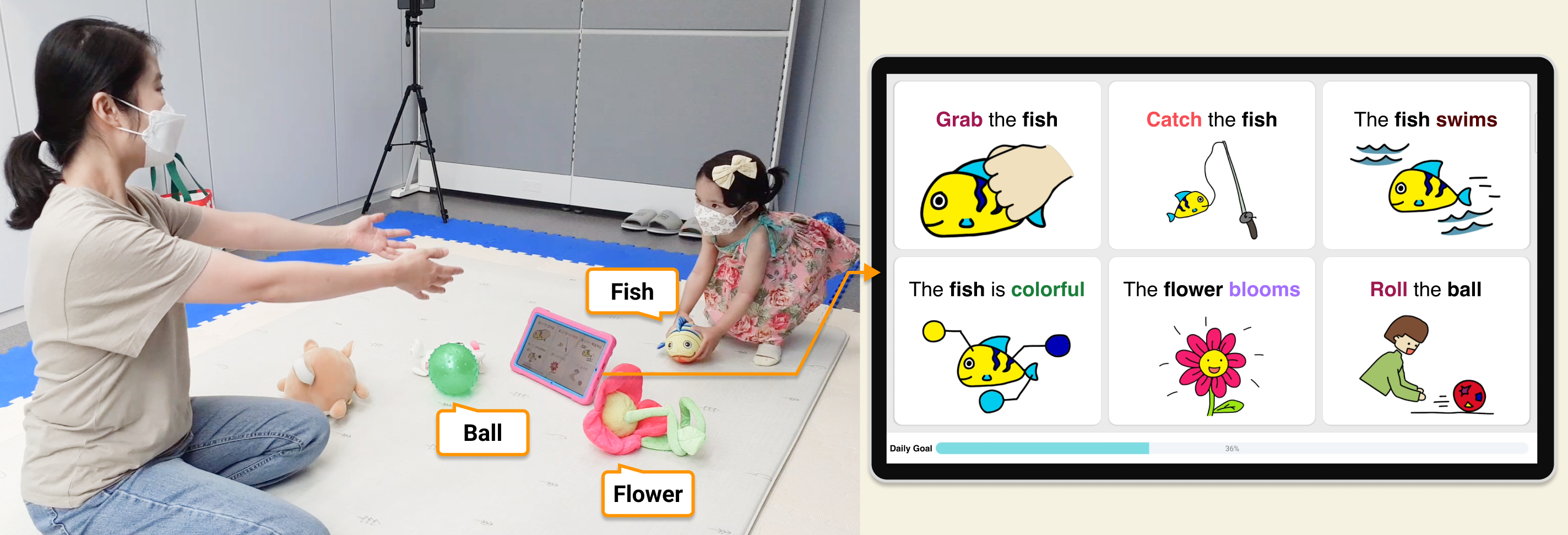}
  \caption{\system{} adapts its guidance to the current play context.}
  \label{fig:teaser}
  \Description{Snapshot of the experiment and the app screenshot. The mother and child are sitting on the floor facing each other. There are tablets, fish, balls, and flowers on the floor between the two. The speech bubble indicates that each object is a fish, a ball, and a flower. The enlarged screen of the tablet in the photo is on the right side of the figure.}
\end{teaserfigure}

\maketitle

\input{sections/1_introduction}
\input{sections/2_background}
\input{sections/3_design}
\input{sections/4_formative}

\input{sections/5_features}

\input{sections/6_prototyping}

\input{sections/7_implementation}

\input{sections/8_evaluation}

\input{sections/9_discussion}

\input{sections/10_conclusion}

\begin{acks}
We first and foremost thank the parents, children, and
experts who have participated, as well as the 
Gwanak-gu, Guro-gu, and Yangcheon-gu
Multicultural Family Support Centers.
We thank the anonymous reviewers for their insights.
We thank Eunjung Oh, members of SNU HCS Lab, Insup Choi,
JaeWon Kim, Jaeyoon Song, Jane Im, Ji Won Yang, 
Taewan Kim, Yoonjeong Cha, 
and Kyusoon Kim for their feedback.
This work was supported by Institute of Information \& communications Technology Planning \& Evaluation (IITP) grant funded by the Korea government (MSIT) (No. 2019-0-01371, Development of brain-inspired AI with human-like intelligence). The corresponding author is Youngki Lee.
\end{acks}

\bibliographystyle{ACM-Reference-Format}
\bibliography{references}

\input{sections/11_appendix}

\end{document}

%% file: sections/1_introduction.tex
\section{Introduction}
Parental language input plays a critical role in the language development of young
children. Termed \textit{language nutrition}, the language that a child encounters 
acts as vital ``nourishment'' for neurodevelopment, especially
at young ages when the foundation for primary language ability is being formed
\cite{zauche2016influence, topping2013parent}. 
Crucially, early linguistic ability
has cascading effects: delayed language development at age 3 has
been found to affect learning not just during childhood but well beyond---into adolescence and even
adulthood \cite{dickinson2011relation, fiester2010early}. 
Hence the quality of parent language and interaction within the short period of early childhood has disproportionate effects on the child's outcome.

However, studies have found that many parents face difficulties in providing
sufficient language nutrition to their children \cite{hart1995meaningful,
golinkoff2019language}. Importantly, parents of low socioeconomic status (SES),
low educational attainment, and diverse cultural backgrounds such as immigrants
and refugees have been found most vulnerable 
\cite{hart1995meaningful, choi2019needs, macleod2020language}. Such cases of low parent
language stimulation are linked with high rates of delayed language development
in their children: for example, the prevalence of language delay among children
of immigrants in South Korea has been measured to be
as high as 70\% \cite{oh2009preliminary, jeong2004study}. This problem has been famously
termed as the \textit{word gap}, referring to the gap in both the quality and
quantity of language that a growing child encounters depending on their family
environment \cite{hart1995meaningful}.

To aid these families, there exist guidance sessions 
administered by trained therapists, for both parents 
(called \textit{parent training}) and children (\textit{speech/language therapy}).
However, the in-person nature of these sessions impose high time and cost burdens. 
Consequently, automated systems that enable at-home guidance are recently being explored 
as complementary solutions to improving parental language. 
To this date, a small number of systems
have been proposed \cite{hwang2014talkbetter, song2016talklime,
huber2019specialtime, gilkerson2017mapping}. These systems analyze speech
throughout parent--child interaction to provide feedback to parents on
their language habits, which include conversational turn patterns, quantity of speech, 
and dialogue sentiment. With preliminary studies, researchers have demonstrated positive
potential for such technology to provide easily accessible and low-cost aid for parents.

Yet a critical limitation of current designs---especially when compared to a human instructor---is that they neglect 
the \textit{context} of interaction.
As an example, the utterances \textit{``That car is fast! Watch out!''} can familiarize
a child with the linguistic entity that is a car, the concept of speed, and 
an idiomatic expression relevant to this setting. However, these words
would be irrelevant and even counterproductive when there is no
fast car around. Context is especially important in the early stages of
language acquisition as associations
between words and the outside world are beginning to take form \cite{brown1958words}.
However, existing systems lack awareness of physical surroundings
altogether as they rely on speech signals.
Moreover, speech is an especially restrictive source of information in settings with 
younger children, who
communicate through body language and underdeveloped spoken language 
\cite{reilly2006growth}. 

In this paper, we explore the design of the first system that aids parent--child interaction
with \textit{contextual guidance} that reflects the play situation.
\system{} composites multimodal information from AI models for 
gaze, scene, and natural language to estimate which objects 
and toys are being jointly focused on during
play---which we call the \textit{joint attention distribution}---and
display relevant phrase cards on a tablet application.
The automated guidance requires no direct manipulation of the
tablet interface, enabling the parent to stay focused on
playing with their child.
Through an evaluation of \system{} with parents--toddler dyads,
we show that the contextual phrase cards can help parents
engage their child with significantly
more \textit{responsive} dialogue than when using static guides.

Key to our work is a interdisciplinary design process targeted towards
a vulnerable population: immigrant families,
whose children are documented to be at
high risk of developing language delay due to deficiencies in
language nutrition.
We pursue user-centered design with immigrant parents, children, and
professional speech language pathologists who specialize in treating culturally and linguistically
diverse families\footnote{All procedures were approved by the Seoul National University IRB.}. 
Through a formative investigation, we find motivation for a
context-aware system that guides parents in providing responsive, 
diverse, and abundant dialogue to young children.
We prototype \system{} with users and experts
to address unique design considerations for context-driven 
automated guidance.
We find and discuss insights
into designing guidance technology for linguistically diverse
families, which is yet underexplored in the HCI literature.

We contribute:
\begin{itemize}
    \item Insights from designing language guidance technology with
    immigrant parents, children, and speech language pathologists.
    \item \system{}, the first contextual language guidance system for parents
    of young children.
    \item An evaluation of our system with parent--child dyads, demonstrating its
    potential to improve the quality of parental language input.
    \item A discussion on key considerations when designing
    contextual guidance technology for linguistic minority families.
\end{itemize}

%% file: sections/2_background.tex
\section{Background and Related Work}
In this section, we introduce the word gap problem in immigrants.
Next, we highlight prior systems for parent--child interaction and second-language 
learning, which jointly motivate our design.
\subsection{Immigrant families and the word gap}
A large body of empirical research has found that the quality of
early parent language affects child cognitive development
\cite{romeo2018beyond, romeo2018language, merz2020socioeconomic},
vocabulary learning
\cite{rowe2008child, cartmill2013quality, rollins2003caregivers},
as well as language outcomes for children with disabilities 
\cite{haebig2013contribution, warren2010maternal}---hence giving 
rise to the term \textit{language nutrition} \cite{zauche2016influence}.
A grim implication of these findings is that
insufficient language exposure may cause some children to
fall behind their peers in cognitive and language development.
Indeed, the \textit{word gap} phenomenon was first documented by Hart \& Risley in
1995 when they observed that children of low-SES hear an average of
30 million fewer words than their high-SES peers, mostly from
their parents---just within the
\textit{first three years} of their lives \cite{hart1995meaningful}. 
This gap in child-directed speech was linked to
the gap in the language abilities of children.

In our work, we target the design process towards
immigrant families in Korea---defined as those with at least
one parent born outside the country---whose vulnerability 
to the word gap problem is well known.
Up to 70\% of young children 
in these families are estimated to exhibit delayed language development \cite{oh2009preliminary, 
jeong2004study, lee2008study}, 
an astoundingly high figure when considering that over 
260,000 second-generation
immigrants are currently estimated to reside in the country \cite{ministry2020status}.
A large part of the problem is attributed to low language
use at home,
as the primary caregiver's unfamiliarity
with the dominant, domestic language acts as a barrier to parent--child interaction
\cite{lee2012language}.
Ideally, active interaction in the caregiver's more comfortable language
can provide the child with sufficient language exposure. 
Unfortunately, this is often not the case; as Section \ref{section:therapist_group_interview} of our
formative investigation will show, factors such as cultural acceptance,
educational concerns, and pressure from family members 
discourage many parents from raising
their children in a bilingual environment.
Finally, a high percentage of immigrant families are financially burdened,
having less means to rely on professional help \cite{ministry2018survey}.

This phenomenon is not unique
to Korea: studies have observed high rates of language
delay and lower amounts of home language stimulation
in various immigrant and refugee populations around the
world \cite{mayo2008off, salam2019communication}.
As a result, there are emerging efforts to design
parent-directed language intervention and guidance programs
tailored to linguistically diverse families 
\cite{becklenberg2021training}.
However, technological aids are yet 
scarce \cite{baralt2020hablame}.
We see a need for further research on both understanding
the use of technology in immigrant families, as 
well as designing supportive artifacts, and 
aim to explore both directions in our work.

\subsection{Automated guidance for parent--child interaction}
Works in the field of automated
language guidance for parents are yet few. 
Within the domains of HCI, CSCW, and ubiquitous computing, researchers have
explored real-time interventions to aid parents 
on interacting with children affected by language disorders. 
These interventions mostly take
on the form of a mobile app, in many cases accompanied 
by other sensing hardware such as wearable microphones.
TalkBetter \cite{hwang2014talkbetter} monitors conversational turns and
triggers a warning when a parent displays detrimental language habits---for
example, interrupting the child while they are talking.
TalkLime \cite{song2016talklime} similarly monitors turns, but uses a
real-time visualization on the parent's mobile screen to
induce behavior change. 
Other works have targeted
specific contexts. 
SpecialTime \cite{huber2019specialtime} is
designed to be used in Parent--Child Interaction Therapy (PCIT),
which requires tracking a parent's performance on a predefined
set of guidelines. By using speech
recognition and sentence classification on parent dialogue,
SpecialTime allows the parent's language to be evaluated at home
instead of clinical locations.
On the commercial side, the predominant tool is LENA
\cite{oller2010automated, gilkerson2017mapping}, which estimates word and turn counts of parents
and their child,
serving as a passive informatics tool.
As part of an effort to reduce the word gap,
the LENA software has been distributed to
low-income parents \cite{wong2020providence}.

Importantly, all prior systems rely on speech processing.
However, audio is a noisy and restrictive signal 
where young children are concerned, as gestures, cries,
and prelinguistic vocalizations prevail
\cite{reilly2006growth, ingram1989first}.
This leads to challenges in both accuracy and specificity of
guidance.
For example, the performance of the LENA system in tracking
child-directed words and conversational turns
has been found to be low 
\cite{cristia2020thorough, mcdonald2021evaluating}. At the same
time, the instructions to parents provided by current systems 
are context-agnostic---
for example, ``speak more words''---while being unable to
recommend \textit{which} words to use in the current situation.
We see an opportunity to expand the quality of 
guidance by incorporating \textit{visual} signals
to form a more comprehensive picture of interaction context.

Our work also builds upon a broader body of system designs for parenting.
While not providing real-time feedback, 
H{\'a}blame Beb{\'e} \cite{baralt2020hablame} is a recent
mobile app that shows promise in teaching parents about 
the importance of 
language nutrition and bilingual home environments.
MAMAS \cite{jo2020mamas} aids mealtime situations with specialized
hardware that tracks the child's eating habits, along with an app for parent
reflection. MOBERO \cite{sonne2016changing} assists parents with children
affected by ADHD by guiding them through morning and bedtime routines.
ParentGuardian \cite{pina2014situ} also supports parents with ADHD children by
detecting moments of high stress, then providing guidance to lower stress levels. 
DyadicMirror \cite{kim2020dyadic}
aids parents' self-awareness by fitting the child with a wearable
``mirror'' device, which display's the parent's face throughout interaction.
The positive behavior changes demonstrated by these systems 
points to the many opportunities that lie in designing automated 
guidance technology for parents.

\subsection{Contextual second language learning tools}
Since the advent of mobile devices, researchers have explored 
a variety of context-aware applications to enhance learning. 
Most of these works have focused on
second language learning in adults. Contexts incorporated range from 
the user's location to nearby objects \cite{lee2019systematic}. 
An early work, MicroMandarin \cite{edge2011micromandarin}, is a mobile
application that takes the user's location and generates flashcards of
context-relevant phrases. For example, in a coffee shop, the user might be
quizzed with a flashcard displaying the Chinese word, ``hazelnut''. Vocabura
\cite{hautasaari2019vocabura} similarly uses location information to recommend
words, but through real-time audio. More recently, Draxler et al.
\cite{draxler2020augmented} proposed an augmented reality (AR) application that
dynamically generates quiz questions based on the location of objects. An
example might be ``the cup is next to the \textit{blank}'' where the user must
select the correct noun.
Combined, these works have demonstrated that dynamic, context-aware systems
are effective educational tools for learning language, which is fundamentally
an associative and contextual process.

\subsection{Summary of related works}
In summary, insufficient language nutrition causes
many children to fall behind their peers from early on;
one especially vulnerable population are children growing
up in linguistically diverse homes.
Automated systems that guide parent language use show potential to 
be an accessible and effective solution for these families.
However, currently explored systems do not consider
contextual information, which contains diverse
word learning cues such as visual and situational 
information.
Meanwhile, many works have found context-aware technology
to be effective tools for second-language learning in adults.
Informed by these ideas,
we aim to design the first contextual language guidance system
for immigrant parents to help them provide effective 
language nutrition to their children.

%% file: sections/3_design.tex
\section{Design process}
\label{sec:design}
\begin{figure}
	\centering
    \includegraphics[width=\columnwidth]{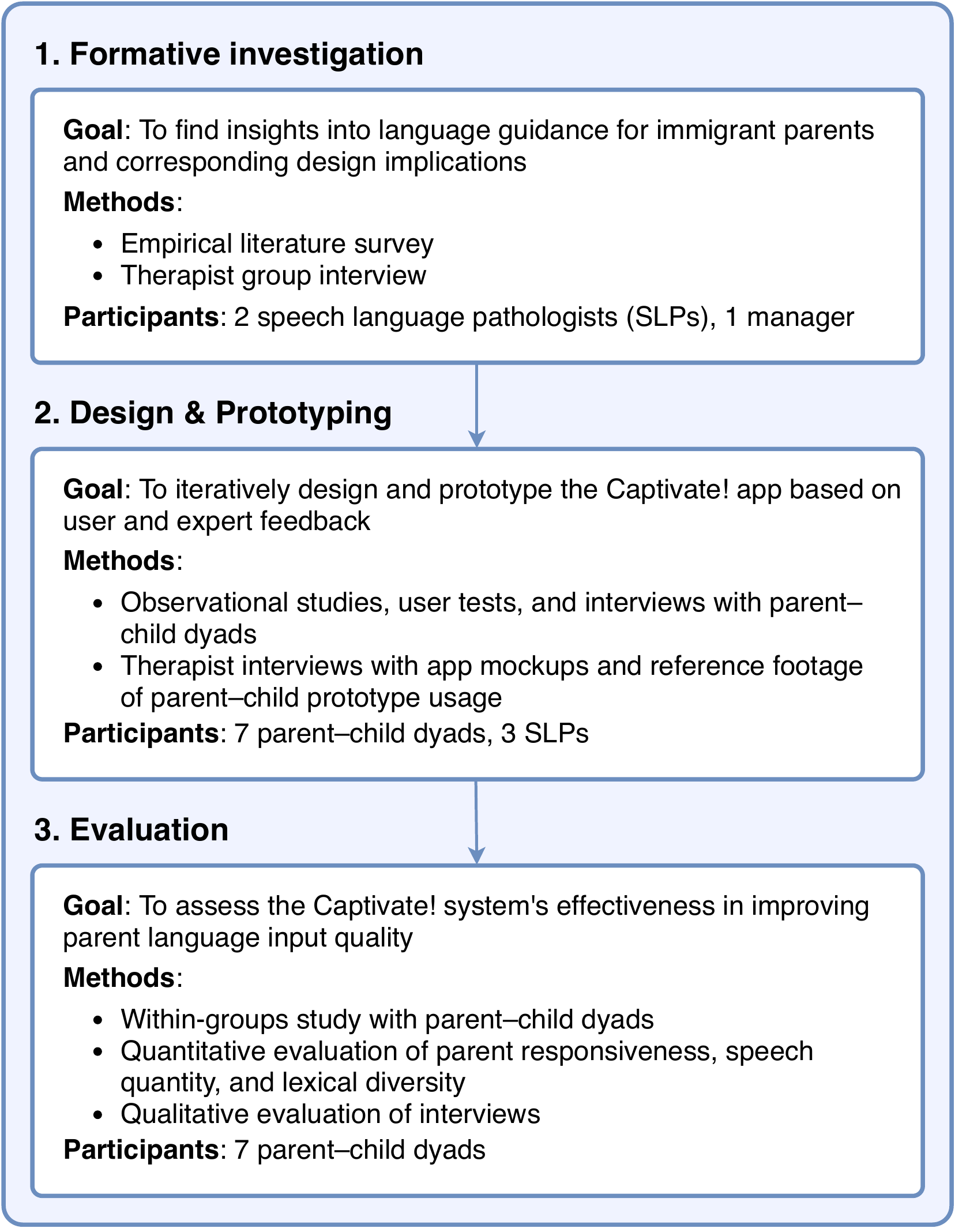}
	\captionof{figure}{Overview of the
	study procedure.}
	\label{fig:study_procedure}
	\Description{Flowchart describing the overview of the study procedure. Details are provided in section 3.}
\end{figure}

To design our system, we structured a user-centered design
process with parents and children of immigrant families, 
and experts specializing in speech/language guidance for 
these households.
Our procedure, shown in Figure \ref{fig:study_procedure}, consisted of multiple design stages
that can be grouped into two.
First, we conducted a formative investigation (Section 
\ref{sec:formative})
where we established goals and considerations for a
technological aid that improves parents' language, specifically in
the context of immigrant families.
Here we explored the empirical literature
on which parent language qualities affect
children's development. Furthermore, we conducted an interview with
language guidance practitioners to understand how we can
design a system that meets the guidance needs of the 
target user population. Combined, these
processes revealed key design implications for our system. 

Second, we designed features that directly address
these implications (Section \ref{sec:features}), 
and iteratively prototyped the resulting system, \system{},
with users and experts (Section \ref{sec:prototyping}).
Our methodology included observational studies and 
usability tests with 
parent--child dyads, interviews with parents,
and interviews with SLPs using low-fidelity mockups and
usage videos. 
We discovered considerations for the novel context-driven
interface that functions without user manipulation,
and improved its usability through user feedback.
We also iterated over the guidance content displayed
to parents, and designed phrases that stimulate
parent language and interaction.
Finally, we improved the accessibility
of the app for parents with low Korean proficiency.

\subsection{Participants}
Parent and expert participants were recruited through
the help of
Multicultural Family Support Centers\footnote{\url{https://www.liveinkorea.kr}}, which direct government
efforts to support immigrant and foreigner families in Korea.
Among the centers' initiatives is a language development
program for children in these families. In the program, 
SLPs employed by the center administer 
speech/language therapy and parent
training. We initially reached out to five centers and
explained our study goals and procedure, leading to three 
centers agreeing to participate. 
It should also be mentioned that previously,
we had tried several traditional channels for recruiting, such
as online advertisements, but were unable to reach immigrant
parents through them.

\subsubsection{Parents and children}
The parents and children who were recruited
are shown in Table \ref{tab:parent_child_participants}. 
All parents were recruited through the three centers
either directly (by center personnel calling or asking
in-person) or indirectly (through a friend or family member
that had been reached).
We note that many parents visit the centers because
they have children participating in the language development 
program, which is mainly offered to children who
show signs of language delay (as judged by the SLPs through
a formal evaluation).
This likely affected the prevalence of language delay in our
child participants, as well as the level of familiarity the 
parents had with language guidance.
Parents of multicultural families (defined as 
those with one parent born outside of 
Korea) or foreigner families (both parents with non-Korean 
nationality) were eligible to participate.
We included children aged 0--5 in our eligibility
criteria for the prototyping process, which we narrowed to
1--3 in the evaluation to add control for child age-related
factors. As our system provides Korean language guidance, 
we verified that all
parents recruited included Korean among the languages spoken to
their children. Otherwise, we did not screen for proficiency 
in Korean.

\subsubsection{Experts}
The expert participants are shown in Table 
\ref{tab:expert_participants}.
With the exception of M1, who was a program manager
in charge of the language development program at a center,
all were SLPs currently employed
at the Multicultural Family Support Centers.
The SLPs specialized in screening children of immigrant families 
for language problems, administering language development sessions, 
and instructing parents. 
M1 had multiple years of experience in
coordinating programs for immigrant and foreigner families.
All three experts participating in the formative investigation
were recruited from the same center; all SLPs of the prototyping
stage were recruited from another (single) center.

\newcolumntype{Y}{>{\centering\arraybackslash}X}
\begin{table}[]
    \caption{List of parents and children who participated
    in the design and evaluation studies (separated by the middle
    line).}
    \label{tab:parent_child_participants}
    \renewcommand{\arraystretch}{1.1}
    \begin{tabularx}{\columnwidth}{l l l 
        >{\hsize=.9\hsize\linewidth=\hsize}Y
        >{\hsize=.9\hsize\linewidth=\hsize}Y
        >{\hsize=1.2\hsize\linewidth=\hsize}Y 
        Y}
    \toprule
    \textbf{ID} & \textbf{P} & \textbf{C} & \textbf{C age (m)} 
        & \textbf{Lang. delay} & \textbf{Prev. country} 
        & \textbf{Yrs in \newline Korea} \\
    \midrule
    P1, C1    & F  & M & 42 & Y  & Taiwan    & 6     \\
    P2, C2    & F  & F & 39 & Y  & China     & 10+   \\
    P3, C3    & F  & M & 34 & Y  & China     & 10+   \\
    P4, C4    & F  & M & 39 & Y  & Cambodia  & 4     \\
    P5, C5    & F  & F & 48 & Y  & China     & 10+   \\
    P6, C6    & F  & M & 48 & N  & China     & 6     \\
    P7, C7    & F  & M & 69 & Y  & China     & 3     \\
    \midrule
    P8, C8    & F  & F & 19 & N   & Vietnam  & 10+   \\
    P9, C9    & F  & F & 50 & N   & Vietnam  & 5     \\
    P10, C10  & F  & F & 31 & Y   & Korea    & 10+   \\
    P11, C11  & F  & F & 41 & N   & Vietnam  & 3     \\
    P12, C12  & F  & M & 20 & N   & Vietnam  & 10+   \\
    P13, C13  & F  & F & 47 & Y   & Vietnam  & 5     \\
    P14, C14  & F  & F & 31 & N   & Vietnam  & 8     \\
    P15, C15  & F  & F & 27 & Y   & China    & 10+ \\
    \bottomrule
    \end{tabularx}
\end{table}

\begin{table}
	\centering		
	\caption{List of experts who participated in the
	design process.
	Experience here refers to the number of years at 
	Multicultural Family Support Centers
	specifically and excludes prior training.
	}
	\label{tab:expert_participants}
	\renewcommand{\arraystretch}{1.1}
    \begin{tabularx}{\columnwidth}{p{15pt} X X X}
    \toprule
    \textbf{ID} & \textbf{Profession} & \textbf{Stage} & \textbf{Experience}\\
    \midrule
    S1 & SLP & Formative  & 1 year \\
    S2 & SLP & Formative & 3 years \\
    M1 & Manager & Formative & 1 year \\
    \midrule
    S3 & SLP & Prototyping & 5 years \\
    S4 & SLP & Prototyping & 1 year \\
    S5 & SLP & Prototyping & 2 years \\
    \bottomrule
    \end{tabularx}
\end{table}

%% file: sections/4_formative.tex
\section{Formative investigation}
\label{sec:formative}
We began our design process with a formative investigation that
explored the following research questions:

\begin{quote}
\begin{itemize}
    \item[\textbf{RQ1.}] \textit{Which} qualities of parental language input should our system aim to improve?
    \item[\textbf{RQ2.}] \textit{How} can our system guide parents 
    into improving these qualities?
    \item[\textbf{RQ3.}] What special design considerations are needed for \textit{immigrant parents}?
\end{itemize}
\end{quote}

Our investigation consisted of two parts.
First, we explored the empirical literature
on the relationship between parent language and child language acquisition,
gaining insights into \textbf{RQ1}.
Second, we conducted a group interview
with experts who specialize in counselling culturally diverse 
families to seek answers to \textbf{RQ2} and \textbf{RQ3}. 
In this section, we share our findings and summarize 
the corresponding design implications.

\subsection{Exploring goals of language guidance}
Language is complex, encompassing everything from
semantics, to grammar, to pragmatics such as nonverbal gestures
\cite{fromkin2018introduction}. We sought to
scope this design space by asking,
``what are the \textit{key qualities} of parent language that 
act as language nutrition, stimulating language growth in 
young children?''
To answer this question, 
we referred to large-scale meta-analyses of empirical studies
that have linked various attributes of parent behaviors to 
the language outcomes of their children 
\cite{zauche2016influence, topping2013parent}.
As a result, we found several key qualities of parental input
that are thought to facilitate language acquisition,
and summarize them below.

\subsubsection{Responsiveness}
Parent responsiveness can be defined as the “prompt, contingent, 
and appropriate reaction to infant behaviors” \cite{mcgillion2013supporting}. 
For example, a responsive parent would promptly respond to 
a child pointing at a car with a relevant comment, such
as “Yes, that is a red car!”. 
In this example, the parent demonstrates
two important responsive behaviors:
\textit{verbal contingency}---responding to the child with
appropriate speech---as well as 
\textit{joint attention}---being focused on the same subject
as the child.
Responsiveness is thought to be one of the most important
qualities of effective parent input as its positive
effects on infant language learning has been 
widely documented
\cite{begus2014infants, camp2010relationship, kasari2014randomized, hoff2006social, laakso1999early, nicely1999mothers}. 
For example, a study has found that
infants learned far more effectively when taught about the objects
they personally pointed at \cite{begus2014infants}; other studies 
have found strong correlations between measurements for parent
responsiveness and child development 
\cite{camp2010relationship, laakso1999early, nicely1999mothers}.

\subsubsection{Lexical diversity}
The diversity of a parent's vocabulary
usage has been linked to language ability in children \cite{pan2005maternal, hoff2002children, huttenlocher2010sources, burchinal2008cumulative, rowe2012longitudinal}. 
Quite intuitively, children exposed to richer words 
have more opportunities to expand the breadth and
depth of their vocabulary understanding.
Some studies have even found that parent lexical diversity
accounts for a significant portion of the learning gap 
between low and high-SES children 
\cite{huttenlocher2010sources, burchinal2008cumulative},
which highlights its importance.
However, it is important to note that simply including
obscure vocabulary words in one's dialogue just for the
purpose of diversification is unlikely to be effective.
Instead, the variety of words that are used
must exhibit responsiveness, encourage the child's 
interaction, and include rich associations between
real-world entities and language \cite{justice2018evidence}.

\subsubsection{Quantity of speech}
More talk by the caregiver predicts
child language outcomes 
\cite{hart1995meaningful, rowe2008child, gilkerson2009power, shneidman2013counts, weisleder2013talking, romeo2018beyond}. 
Specifically, studies have linked the volume of parent
words to the future vocabulary of the child 
\cite{rowe2012longitudinal} and 
cognitive development \cite{romeo2018beyond}.
However, as in the case of lexical diversity, simply speaking without
thought or leaving the TV on
is unlikely to be effective language input.
Current research postulates that words
\textit{directed} to the child are the primary stimuli behind
language learning:
speech directed to children (but not overheard speech) 
predicts their future vocabulary size 
and is correlated with the processing speed of familiar vocabulary 
to around 30 months \cite{shneidman2013counts, weisleder2013talking}. 
In other words, active language use 
that, once again, exhibits responsiveness and engages the child should
be a goal of parent language.

\subsubsection{Notes}
It is worth noting that literature on child-directed 
speech has traditionally 
described acoustic features such as slower speaking rate, exaggerated intonational contour, and clear pronunciation as some of 
the key features, 
but the direct effects of these features on 
language outcome have been debated 
\cite{liu2003association, haake2014slower, cristia2014hyperarticulation}. 
Importantly, many qualities of parent language are 
intercorrelated: for example, parents who are responsive
are likely to display more positive affect
\cite{zauche2016influence}.
We therefore hypothesize that targeting our guidance 
towards the three key qualities above can also 
yield general improvements
in the quality of parental input.

\subsection{Therapist group interview}
\label{section:therapist_group_interview}
Next, we sought to supplement the insights from the literature 
with the expertise of practitioners.
Specifically, we looked for practical guidance techniques
that can be adapted to a mobile app-based setting, 
as well as considerations when designing for
immigrant parents and children.
We conducted a 90-minute, semi-structured group interview with 
two SLPs (S1, S2) from a 
Multicultural Family Support Center and the
manager (M1) who led its language development program.
We transcribed the interview and applied 
iterative coding to find the following themes:

\subsubsection{There is a need for accessible guidance for parents}
The SLPs shared that many immigrant parents rely on professional therapist 
sessions for their child's language stimulation. 
\textit{``At my previous institution, the mothers mostly had occupations \dots so
they were more dependent on professional language education sessions, 
but the education did not continue at home. So I often thought about
how to extend the education [to home settings]---but should I 
call it their personal lives?
It felt like too much to ask to intrude and tell them, `you must act like this [at home]'...
they signed up for these sessions because they lack 
Korean skills, and they have jobs... so asking them to do too much can be very 
burdensome for them''} (S1).
However, the in-person nature of therapist-led sessions limits their frequency; 
moreover, the COVID-19
pandemic prevented these face-to-face sessions altogether,
leading to a gap in the children's learning.
\textit{``If the app is developed, [parents] can use that. I think it would 
be good if there was a way to provide stimulation at home. Because,
language development [classes] are only twice a week.''} (M1).

\subsubsection{The language barrier often prohibits interaction}
A dominant theme throughout the interview was the difficulties
many parents experience in interacting with their children when
they themselves are uncomfortable in the Korean language.
\textit{``Many parents of young children shared with me that they
find it difficult to linguistically stimulate their child, or
to get them to follow directions''} (S1).
\textit{``There are also parents unable to
read Korean \dots they express regret
that they cannot read a book for their child''} (M1).
This suggests a need for guidance
that helps these parents in providing richer Korean language input
while interacting with the child. At the same
time, it demonstrates the importance of \textit{accessible}
design that can be used even by parents unfamiliar with
the target language, while
also being useful to parents more fluent in it.

Importantly, the SLPs reported a lack of bilingual interaction
in households despite the low Korean proficiency:
\textit{``Of the five children that I'm currently treating, only
one family has a positive attitude towards a bilingual environment.''} (S1);
\textit{``It's about the same for me.''} (S2).
The primary reason was the belief that 
such environments would induce \textit{subtractive bilingualism} \cite{lambert1974culture}, where the
child learns the parent's language at the expense
of Korean, which may disadvantage them socioeconomically.
\textit{``There are many cases when the
primary caregiver, the grandparents, or acquaintances view
[bilingual households] negatively''} (S1); \textit{``It may be a negative
effect of focusing on Korean language development since [the child]
is going to grow up in Korea''} (M1). We fully acknowledge the need for
societies to accept and foster linguistically diverse children, but given the status quo, we scope our design towards 
monolingual (Korean) language stimulation guidance.
We later discuss potential ways to extend guidance technology to
multilingual households.

\subsubsection{There are diverse techniques for language stimulation}
When asked about the most effective guidance technique, 
the SLPs expressed that there is no single answer.
\textit{``Although a therapist may be able to judge which 
stimulation techniques are suitable for specific children, 
we cannot say which techniques are more important in general. 
Because, it really differs by the child's case.''} (S1).
For example, some children have difficulties in pronunciation;
their parents would benefit from lessons in making correct
mouth shapes. Meanwhile, other children require more exposure
to language---in these cases, many of the parent qualities
listed above are targeted. Some techniques include the 
use of situation-relevant words (\textit{``it is good [for language stimulation]
to repeatedly expose the child to words related to the people and things they
encounter in their daily lives''} (S2)), or expansions (\textit{``if the
child says `ball', we can expand it to `large ball' or `soccer ball'
to add a semantic element.''} (S2)). 

Concerning parent language training, one of the SLPs mentioned
their experience with \textit{triadic guidance}, 
where an expert gives real-time
feedback as the parent interacts with their child.
\textit{``[In these sessions] I watched the child and mother
play, and instructed to her, 
`you should provide stimulation like this' \dots 
when the parents watched me and also listened to my feedback,
it was definitely helpful for them in improving with
areas that they could not approach at home.''} (S1).
While triadic guidance has
been found to be effective in the literature 
\cite{roberts2011effectiveness}, in practice the technique is not
often used for immigrant parents as many of them feel uncomfortable
speaking to their child under observation. 
Hence, we saw a valuable opportunity
for a mobile application to act as a unobtrusive
triadic guidance provider, enabling the parent to learn from
active feedback rather than passive guidelines.

\begin{table*}[ht]
\caption{Summary of formative insights, corresponding design implications, and the final features designed to address each.}
\label{tab:formative_insights}
\renewcommand{\arraystretch}{1.3}
\begin{tabularx}{\textwidth}{
    >{\hsize=1.2\hsize\linewidth=\hsize}X
    >{\hsize=1.2\hsize\linewidth=\hsize}X
    >{\hsize=0.6\hsize\linewidth=\hsize}X
}
\toprule
\textbf{Formative insight} & \textbf{Design implication} & \textbf{Designed feature} \\
\midrule

There is a need for accessible, at-home
language guidance tools for immigrant parents. &
A mobile app can function as a convenient, unobtrusive, and active guidance provider. &
Tablet app form \\

Responsiveness is a key quality in stimulating
child language development. &
The app should aid the parent in responding
to their child's actions and interests---without diverting attention
from the main interaction. &
Joint-attention-aware guidance\\

Using language that is
active, diverse, yet also relevant and engaging is another
key quality. &
The app should stimulate the parent into
speaking actively and diversely about the child's focus. &
Phrase cards with target words (joint attention-aware)\\

Immigrant parents have varying levels
of proficiency in the domestic language. &
The app should be accessible to 
parents with low language proficiency,
while maintaining usefulness to more fluent
users.
&
Drawings \& read-aloud function \\
\bottomrule
\end{tabularx}
\end{table*}

\subsection{Design implications}
Combined, the investigations revealed a need for
language guidance technology for immigrant parents, as the
language barrier and social factors hinder rich language
input to their children. 
The language stimulation techniques shared by the SLPs 
aligned with our formative findings on the 
importance of exposing children to
responsive, diverse, and active language; they
also provided key ideas
for mobile app-based parent language guidance.

We summarize the key insights from the formative investigation 
in Table \ref{tab:formative_insights},
and map them to design implications for our system. In the next
section, we address these considerations through the design
of \system{}.

%% file: sections/5_features.tex
\section{System features}
\label{sec:features}
\begin{figure}[t]
    \centering
    \includegraphics[width=\columnwidth]{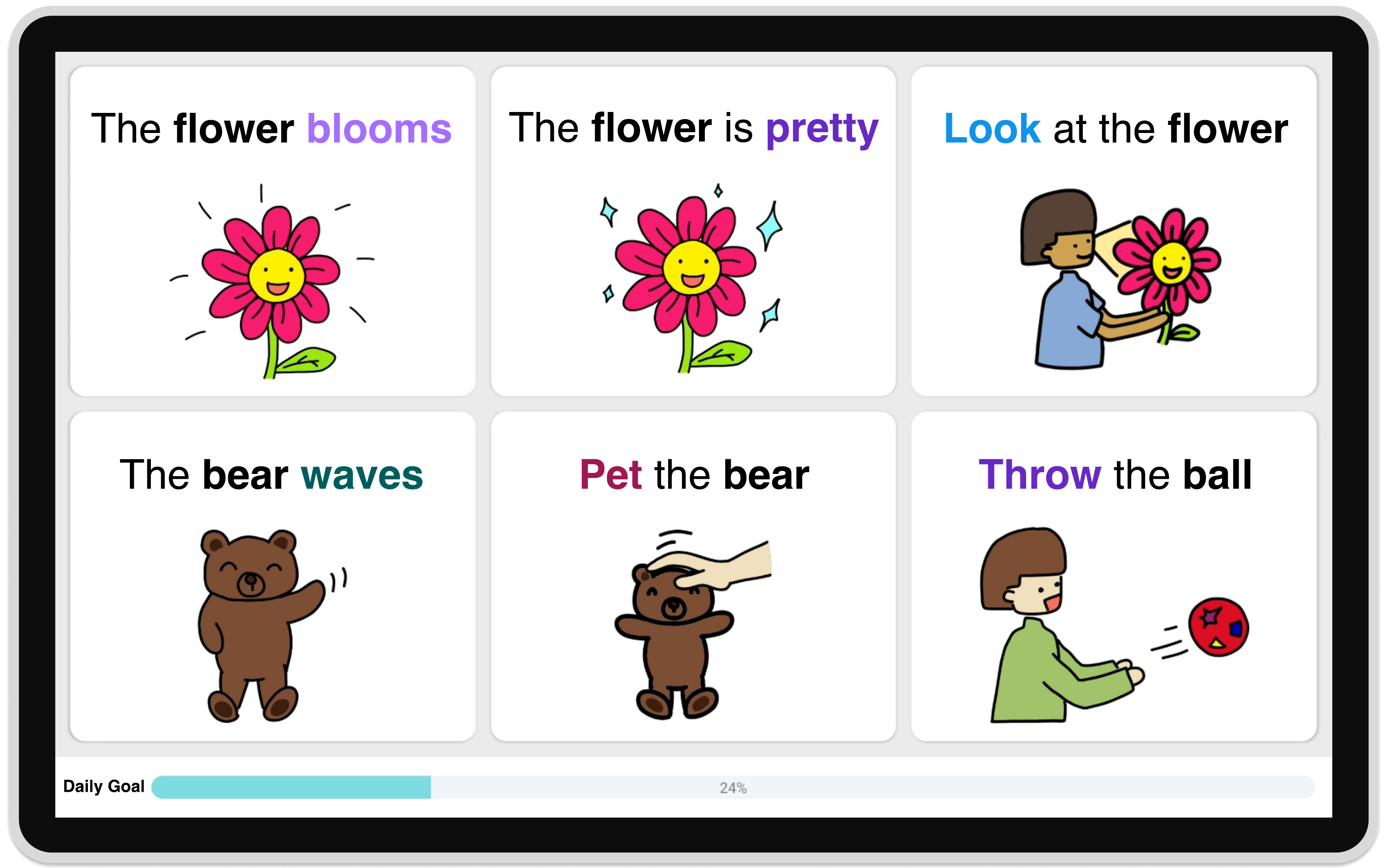}
    \caption{The main interface consists of six 
             phrase cards arranged in a grid, and a progress bar underneath
             that fills as the parent uses more target words (colored).
             The shown cards are dynamically chosen based on the current
             interaction context.}
    \label{fig:app_ui}
    \Description{Application interface with six phrase cards and a progress bar underneath. Phrase cards are arranged in a grid, and each card contains a phrase and a drawing depicting the phrase.}
\end{figure}

\begin{figure}[ht]
    \centering
    \includegraphics[width=1.0\columnwidth]{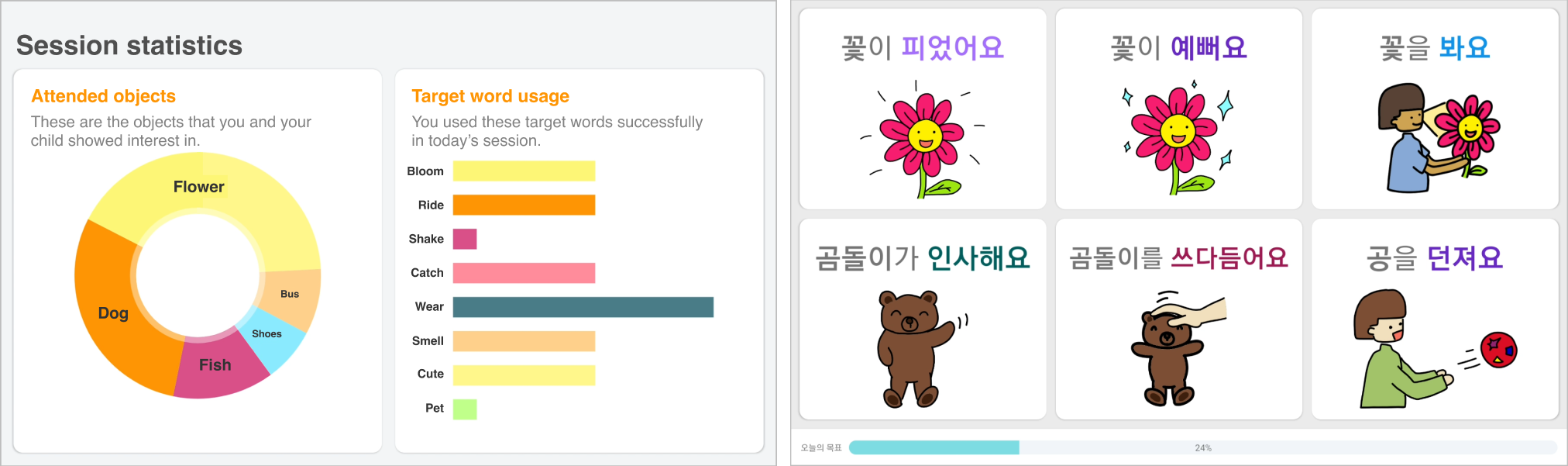}
    \caption{
             (Left) Session statistics based on contextual
             tracking.
             (Right) The original Korean interface used in the study.}
    \label{fig:app_supplementary}
    \Description{A grid of two screenshots. The screenshot on the left has the title "Session statics" at the top of the screen, followed by a pie chart and a horizontal bar chart. The screenshot on the right shows the same interface as figure 3 but the phrases are replaced by Korean.}
\end{figure}

Based on the formative implications, we designed \system{} 
(Figures \ref{fig:app_ui}, \ref{fig:app_supplementary}), 
a tablet application that provides language guidance to parents
throughout play.
The UI consists of several phrase cards;
each card contains a phrase with a target word related to a 
specific object, such as a flower or teddy bear.
\system{}'s defining feature is the real-time, 
\textit{contextual recommendation} of cards that reflects 
the focus of the child and parent. 
For example, upon attending to a flower toy,
\system{} will display phrases corresponding
to flowers; the number of flower-related
cards will increase in proportion as more
attention is focused on it by the parent
and child. No direct manipulation of the app is necessary,
and the tablet is intended to lay statically on the floor.

In this section, we introduce the core system features. 
In Section \ref{sec:prototyping}, we further describe how these
features were concretized through user and expert feedback.

\subsection{Joint attention-aware guidance}
Our formative investigation revealed the importance of
responsive parent language input. 
However, facilitating responsiveness with a mobile app is challenging for
several reasons.
First, responsiveness requires being aware of the current interests
and communication gestures of the child, as well as the surrounding
environment, such as the toys that are being played with. 
For example, if the child shows interest in a doll, 
a responsive parent would
quickly respond with doll-related dialogue. 
However, current mobile apps are not able to
understand situations occurring outside the screen. 
Second, responsiveness requires the parent to be focused on
the child, not a mobile application. Thus, an app that facilitates
responsive behavior in real-time must be able to do so while
detracting minimal attention from the parent. 

To address these challenges, we conceived of guidance
that automatically adapts to \textit{joint attention} between a
parent and child. Joint attention is
a broad concept encompassing a shared focus on an object or person \cite{moore2014joint}. 
By encouraging
language that is relevant to the targets of joint attention, 
we can facilitate responsiveness.
However, yet another challenge lies in that 
young children's attention spans 
are short and dynamic.
Within seconds, the child---and a responsive parent---may shift their attention
from playing catch, to riding a bike, to eating.
Consequently, providing guidance relevant to only a 
single attended object at a given
moment in time would lead to frantic changes in the user interface.
We therefore modeled attention as being distributed across several objects,
which we call the \textit{joint attention distribution}.
The weight of each object in the distribution determines 
the corresponding amount of guidance to be shown.

Importantly, attention is not conveyed through a single action; 
speech, body language such as pointing, and gaze can all 
signify one's focus \cite{birmingham2009human}. 
Based on this observation, we designed 
a system that infers targets of joint attention based 
on \textit{multimodal} cues---specifically gaze and speech, 
which are among its most prominent signals.
An illustration of the mechanism is shown in 
Figure \ref{fig:joint_attention}. 
The targets of joint attention are signified 
both through the parent and child's gaze,
as well as the parent's dialogue 
\textit{"throw the ball!"}. 
Combined with the joint attention distribution model,
we can ensure that as focus is shifted to the ball, 
there is a corresponding increase in the number of 
ball-related guidance cards on the screen. 

\subsection{Phrase cards with target words}
To stimulate active and diverse parent language,
we selected \textit{phrase cards} as the form of
guidance. Phrase cards are widely used by parents
and SLPs as both a reference
and play object when interacting with children.
We conceived a simple form of cards containing
\textit{target words} relevant to an object:
for example, a target word for ``ball'' can be
''throw''; one for ''flower'' can be ''colorful''.
In effect, the target words can help the parents 
apply productive language features such as 
situation-relevant words
or expansions.
We initially referred to a curated list of words
for language-learning children to select target words for 
common objects and toys \cite{chang2013study}.
However, we progressively iterated through different target word
selections and phrase structures through user and expert 
feedback---a process which we detail in Section 
\ref{subsec:phrase_prototyping}.
Additionally, we added motivation for parents to use 
the target words with a \textit{daily goal} progress bar that 
increments as their usage is detected.

%% file: sections/6_prototyping.tex
\section{Prototyping} \label{sec:prototyping}
We iteratively prototyped \system{} with parent--child dyads and SLPs
to improve its usability and effectiveness.
A thorough prototyping process was especially 
important given the novelty of the context-driven guidance interface,
as well as the limited guidelines on designing technological
language guidance for the target user population.

\subsection{Methodology}
The participants were 7 parent--child dyads from multicultural 
homes (P1, C1, \ldots, P7, C7)
and 3 SLPs (S1, S2, S3), as shown in Tables 
\ref{tab:parent_child_participants} and
\ref{tab:expert_participants}.
Specifically, we conducted two groups of studies: 
(1) observational studies, usability tests, and interviews with parent--child dyads; and 
(2) expert interviews with low-fidelity mockups and prototype usage videos.
Due to the COVID-19 pandemic, we conducted two of our parent feedback sessions remotely over Zoom. The other five sessions consisted of on-site studies where a parent brought their child to a playroom
to use the prototype. Each play session lasted 
around 60 minutes, with half the time spent playing with the
child naturally (without using the prototype). We left the
room but recorded the play footage, which we reviewed afterwards.
An interview was conducted after each session\footnote{We share
the guiding interview questions in the Supplementary section.}.
With the SLPs, we conducted 60-minute 
semi-structured interviews at two different stages of design:
once with low-fidelity mockups (with S3) and once using reference
videos of dyads using the prototype in a play situation (with S4, S5). 
We continuously iterated through improved system designs throughout the process.

\begin{figure*}
    \centering
    \includegraphics[width=\textwidth]{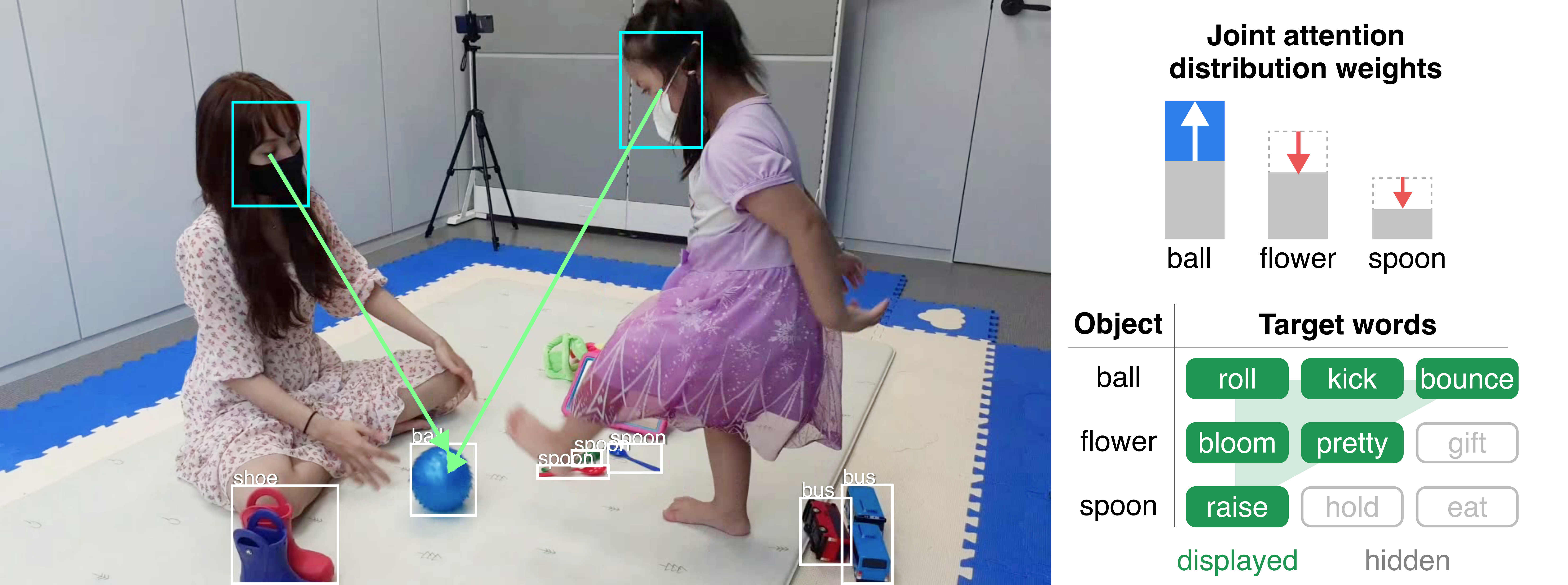}
    \caption{Example of how gaze cues can be used to track
             joint attention. Outputs are
             actual \system{} predictions. Green lines show the 
             model-predicted gaze targets of each person. As focus 
             shifts to the ball, its weight in the joint attention 
             distribution increases while the weights of other objects 
             (such as the previously attended flower) decrease. 
             Finally, the displayed target words are adjusted
             in proportion to the object weights.
             }
    \label{fig:joint_attention}
    \Description{The figure contains one photo and two charts. The photo on the left shows the child and mother playing. The mother is sitting on the floor, and the child is standing and kicking the ball. Blue bounding boxes are on their faces, and arrows indicate the mother and child's gaze as described in the caption. White bounding boxes are on each object, and labels of the objects are on the top-left of each bounding box. A bar chart and a table diagram on the right side of the photo show joint attention distribution weight change and selection of the target words, respectively.}
\end{figure*}

\subsection{Improving the context-driven interface}
\subsubsection{Attention sensitivity}
As the app functioned without direct manipulation from the user, 
it became apparent that \textit{how quickly} the interface
responded to shifts in joint attention was a critical factor
in usability.
We termed this \textit{attention sensitivity},
and explored the answer through observing parents
use the prototype with varying degrees of sensitivity.
When the recommended phrases updated too quickly
to attentional changes, parents
commented that they needed more time to read the phrases, as
playing with the child was already consuming much of their attention. 
However, when the app was overly insensitive, 
many parents spoke about the ``slowness'' of the
app: \textit{``It was difficult to use the sentences 
because the [relevant] cards appeared so slowly''} (P5).
We asked the SLPs for their opinions on the optimal sensitivity:
S3 suggested that roughly 30 seconds would be appropriate.
Based on these insights, we modeled the sensitivity such that
a shift in attention to an object triggers a relevant phrase 
to appear quickly (within 10--30 seconds).
However, additional phrases appear at a slower rate.
The nonlinear sensitivity enabled the interface to be perceived 
as responsive while not overwhelming the user with UI changes.

\subsubsection{Content diversity}
On the other hand, our observations also revealed periods of
prolonged attention towards the same object. 
In this case, there would be
no changes in the recommended phrases, which was problematic
to our goal of stimulating diverse dialogue. 
Parents who evaluated prototypes in this state pointed out that the same sentences continue to be displayed on the screen. \textit{"It's too much of the same sentence. I want the app to show me some other sentences"} (P6). 
Consequently, we designed a mechanism for 
ensuring \textit{content diversity}.
We designed our system to update the phrase of an object-related
card in two cases: first, if the parent speaks
the phrase multiple times, signifying that the child has 
heard it sufficiently; and second, if a phrase had been displayed for a long time, implying that the parent had enough time to read 
and consider it.
For example, once a parent uses the phrase ``throw the ball''
multiple times, the card would update itself to now display 
the phrase ``roll the ball''.

\subsection{Designing effective phrases}
\label{subsec:phrase_prototyping}
Another focus of iteration was in designing phrases
that effectively stimulates language use.
When asked about the difficulties they faced in speaking
actively, many parents reflected that they often could not
quickly think of, and lead the child into, new \textit{situations} 
where different words can be applied.
\textit{``I don't know Korean well, so [I just say] 
just throw it! Throw it! \dots The only word I could come up 
with [the ball] was ``throw''.} (P6). We therefore selected
target words for each object that could both induce and be adapted in 
a wide variety of play situations, such as ``\textit{roll} (ball)''
and ``\textit{bounce} (ball)''. Additional considerations were
made regarding their presentation.
A basic approach that we began with
was displaying just a list of target words.
However, all SLPs thought that it would be difficult for parents to
construct sentences on the fly when just given a single word.
\textit{"It would be helpful for us [SLPs] to reference 
a list of words, but for people who find it difficult to 
make sentences, simple phrases would be better"} (S3). 
Their opinions were mixed on providing complex sentences
such as ``The ball rolled down the hill.'', which 
includes rich language but is also more difficult to adapt 
in play situations.
Based on these comments, we created 
two-word phrases composed of one object and one target word, 
e.g., ``the dog barks''\footnote{
Note that the English-translated phrases are longer
than two words; however, the idea of combining a subject and
target word to form a simple phrase still applies.}. 
The simple form enables easy extensibility, such as with 
expansions (``the \textit{brown} dog barks'') or onomatopoeia
(``the dog barks, \textit{ruff ruff!}'').

\subsection{Supporting low language proficiency}
Nearly all parents and SLPs emphasized the need to
support parents
with low Korean proficiency. P4 shared that 
\textit{"many parents 
of multicultural families in Korea are worried that 
their incorrect pronunciation will have a negative 
effect on their children"} (P4). Based on the feedback, 
we added 
a ``press to read aloud'' function 
and drawings (drawn by the second author) 
to each card to aid in understanding the 
pronunciation and meaning
of guidance phrases. Subsequently,
we observed many parents use the feature to practice
their pronunciation of each phrase while speaking
it to the child, and received 
enthusiastic feedback.
Additionally, many of the
toddlers showed interest in the feature, often repeating 
each phrase after pressing the cards by themselves.

%% file: sections/7_implementation.tex
\section{Implementation}
\begin{figure*}[t]
    \centering
    \includegraphics[width=0.9\textwidth]{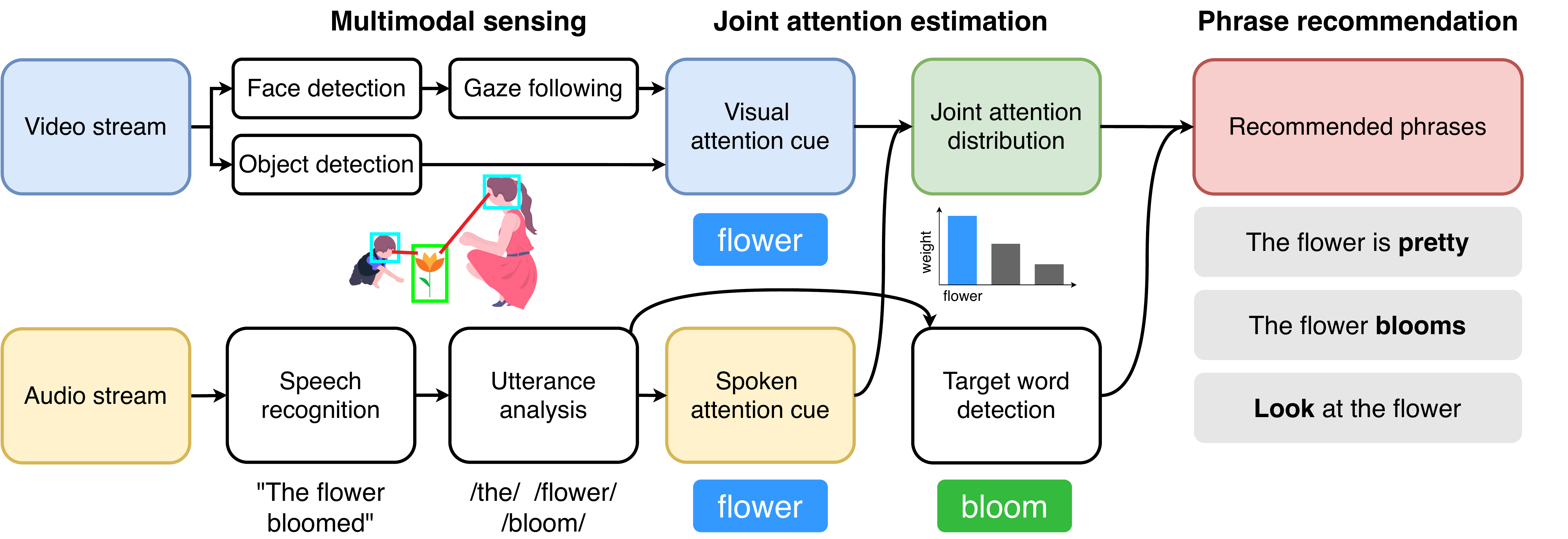}
    \caption{Overview of the processing pipeline.}
    \label{fig:implementation_pipeline}
    \Description{Overview of the processing pipeline. Details are provided in the main text. }
\end{figure*}

To track joint attention, \system{} analyzes real-time 
video and audio of parent--child interaction.
Three tripod-mounted smartphones at different angles 
stream data to a remote server, which performs the
contextual analysis and subsequently sends the relevant phrases 
to the tablet application. 
The app is implemented on Android and communicates
with the server via WebSockets. Video and audio streaming are 
handled through the RTMP protocol. 
Figure \ref{fig:implementation_pipeline} shows an overview
of the main processing pipeline of the remote server, which
analyzes interaction cues, estimates the joint attention distribution,
and selects the relevant phrases.
We explain these in more detail below.

\subsection{Multimodal joint attention sensing}
\system{} extracts attention-related cues independently from
video and audio and use a combination of these
features to estimate the final joint attention distribution.

\subsubsection{Visual attention cues}
As we had selected gaze as the main cue of visual joint attention,
we utilize a recent Gaze Following
\cite{chong2020detecting} neural network to estimate the target
gaze coordinates of both the parent and child's gaze within the scene.
The model requires facial bounding boxes as input, 
which is obtained with a pretrained 
RetinaFace \cite{deng2020retinaface} network.
The system also needs to be aware of \textit{which object}
is located at the attended coordinates.
Therefore, in parallel, we perform object detection on the scene 
using a Faster R-CNN \cite{ren2015faster} model, which
was fine-tuned with labels of 11 toys that were used throughout
the study.
Combined, the pipeline produces estimations of which toys
are being looked at by each individual in each frame. 

\subsubsection{Spoken attention cues}
In parallel, we process the audio stream to find 
spoken attention cues.
Specifically, we detect occurrences of object names in the dialogue
using speech recognition.
For this purpose, we use Google's Cloud Speech-to-Text Service\footnote{https://cloud.google.com/speech/
} to obtain real-time transcriptions of parent speech. 
Next, we perform morpheme analysis and lemmatization on each transcribed
utterance. This serves two purposes: first, the same noun can be modified
in many forms in the Korean language; and second, we can normalize
usage of modified verb and adjective usages as well, such as in the
case of ``the dog ran/is running/will run/runs'', to detect usage
of the target word \textit{``run''}. The latter output is not 
used in joint attention sensing, but rather for the
phrase recommendation stage of the larger pipeline. We use the
KHAiii library\footnote{https://github.com/kakao/khaiii} for analysis.

\subsection{Joint attention distribution}
\label{sec:implementation_joint_attention_distribution}
We use the two attentional cue outputs to estimate a
joint attention distribution, which weighs each toy by
the amount of focus that is currently being placed on it.
We model the distribution
as a temporally dependent set of weights that updates as new
cues are detected. In other words, when there is no
new attentional information---such as when both people are not
looking at a toy---the distribution remains unchanged.
However, if cues are detected, such as when the child's gaze
falls on a flower toy, or if the parent speaks the word ``flower'',
we update the distribution weights with the following policy:
\begin{equation}\label{joint_attention_equation}
    w_{\theta, t} =
    \begin{cases}
        w_{\theta, t-1} + \alpha & \text{if visual cue}\\
        w_{\theta, t-1} + \beta & \text{if spoken cue}\\
    \end{cases}
\end{equation}
where $w_{\theta, t}$ denotes the weight of object $\theta$ at the current
time step $t$ and $\alpha, \beta$ are tunable constants. After updating $w_\theta$, we normalize the distribution to reflect the fact that the
relative weights of all other objects are decreased. This policy
has the effect of modeling attention sensitivity nonlinearly, as shown in Figure \ref{fig:impl_distribution_update}. 

\subsection{Phrase recommendation}
Finally, relevant phrases for each object are chosen and
sent to the tablet app for display. Six cards are displayed
at any given moment, although we note that this number can
be changed. The proportion of phrases 
for a given object reflects its weight in the joint attention
distribution; thus phrases for objects with the highest current
weights are displayed while the others are hidden. To ensure
content diversity, we replace a phrase for an object
in two cases: (1) if the parent uses its target word
$N$ times, and (2) if the card has been on display for over $T$
seconds, where we chose $N=2$ and $T=120$. We prepared six
candidate phrases in total for each object.

%% file: sections/8_evaluation.tex
\section{Evaluation}
We conducted a within-subjects study across two conditions
with $N=7$ participant dyads (parents: 7F; children: 6F, 1M)
to evaluate our final system. We explored the effects
of contextual guidance by comparing \system{} against traditional paper 
phrase cards with identical content.
We note that the selected phrases and card design were already
the result of iterative design and thereby expected to stimulate
parent language use.
However, we used the cards as the control condition to better
view the isolated effects of automated contextual 
recommendations in parent--child interaction.

\subsection{Method}
\label{sec:evaluation_procedure}
\subsubsection{Participants}
We recruited eight participant dyads from Multicultural
Family Support Centers in Korea as described in
Section \ref{sec:design}. There was no overlap in participants
between the prototyping and evaluation procedures.
We later excluded one participant (P9) as her child threw an
extended tantrum during the experiments.
As shown in Table \ref{tab:parent_child_participants}, 
the children were aged 1--3 
(19--47 months; mean = 30.85, SD = 10.30).
Except for one Korean-born parent, all were immigrants to
the country. We did not screen for proficiency
in Korean, leading to wide variety in familiarity with
the language.

\subsubsection{Procedure}
Each parent brought their child to a university facility.
We first introduced the study background and the importance
of active and responsive interactions with children.
Next, the parents and child were led into a playroom 
furnished with a mat, toys, and three cameras set up
on tripods. The parents went through two 30-minute play sessions
with their child in the room using the toys provided, once
with the \system{} system, and once with paper cards.
The order in which these tools were provided was counterbalanced
through alternation, with the first session being determined
randomly. Also, as to minimize bias, only the tool to be used was 
introduced prior to each play session. The experimenter then left
the room to ensure privacy. Finally,
a 10--15 minute exit interview was conducted after the second
session.

\subsubsection{Metrics and coding procedure}
We transcribed and coded a 10-minute segment (beginning from minute 5)
from each session
to analyze several metrics related to our system goals. First,
we quantified the responsiveness of parent language by
coding \textit{verbal responses to the child's focus of attention
or communication acts}. Here we followed the two-step coding procedure
of Haebig et al. \cite{haebig2013contribution} where child engagement
and communication acts were first coded, and parent responses within
these sections were coded thereafter\footnote{More details
on our coding procedure and interrater reliability are in 
Appendix \ref{sec:appendix_coding}.}. 
Next, we evaluated the activeness of parent language
by counting the \textit{number of child-directed words} spoken by the parent.
Third, we evaluated lexical diversity by measuring the
\textit{type-token ratio}, or the number of unique words divided
by the number of total words spoken.
Finally, we also analyzed the accuracy of our system's
contextual recommendations within the segments. 
We share our results below.
\begin{figure}
    \centering
    \includegraphics[width=0.9\columnwidth]{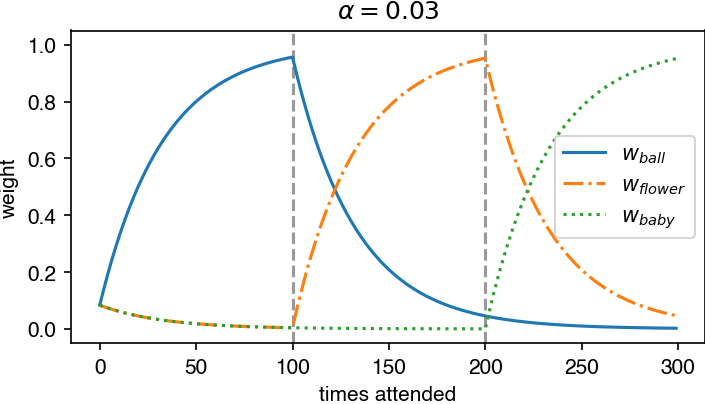}
    \caption{Visualization of joint attention distribution weight updates based on Equation \ref{joint_attention_equation}.
             Dashed vertical lines indicate shifts in attention to a flower and baby toy respectively. Weights are updated quickly upon attentional changes, but the rate slows down as attention is sustained.}
    \label{fig:impl_distribution_update}
    \Description{Line graph showing the joint attention distribution weight updates. The X-axis is time-attached and the Y-axis is weight. It shows that the weight of balls, flowers, and babies increases sequentially.}
\end{figure}

\begin{figure}
    \centering
    \includegraphics[width=0.85\columnwidth]{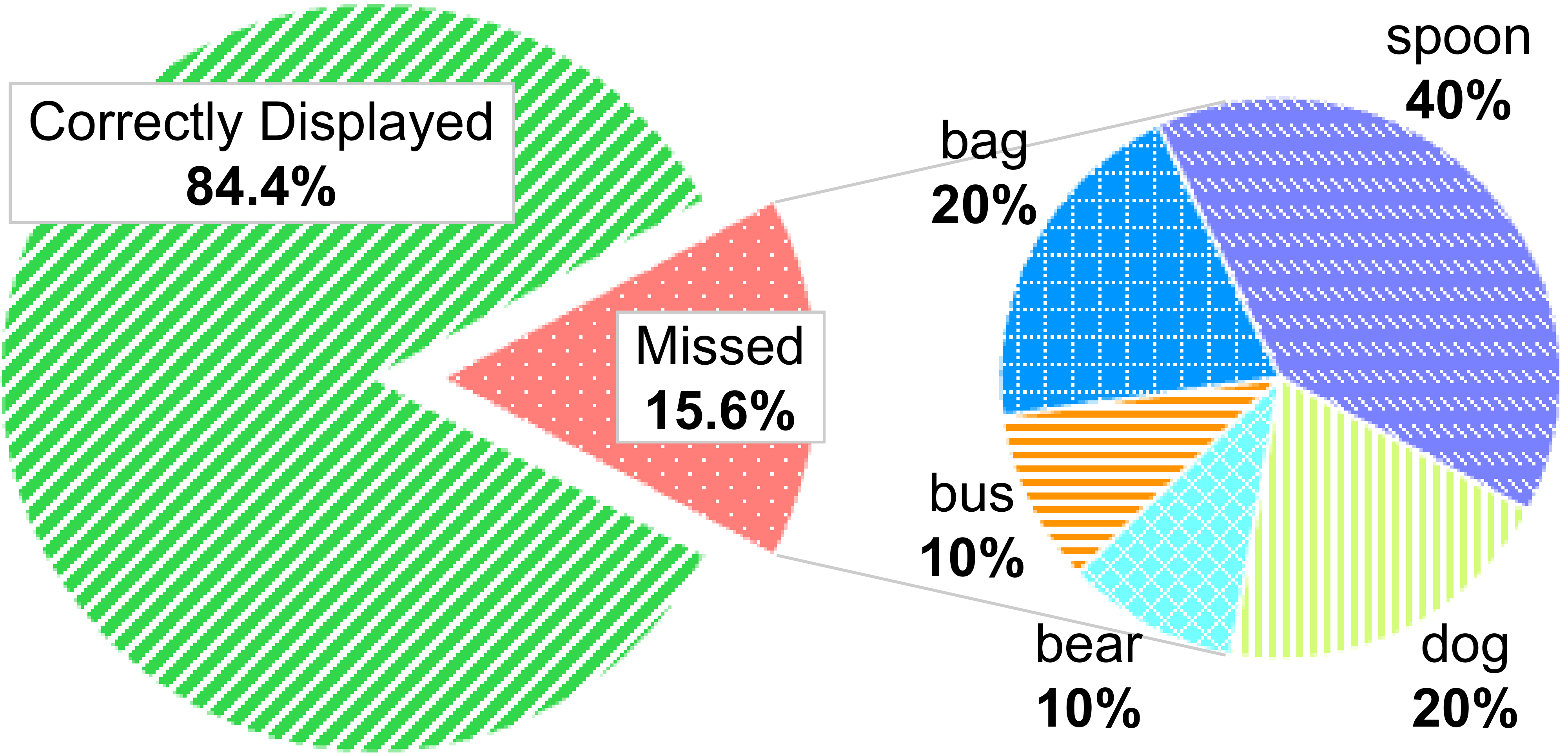}
    \caption{System accuracy evaluation. Within 84.4\% of 
    sampled minutes,
    there was at least one displayed card relevant 
    to the primary target
    of joint attention. The right pie chart shows a breakdown
    of missed objects.}
    \label{fig:results_system}
    \Description{Pie of pie chart showing the system accuracy. The left pie chart shows the ratio of correct and missed, and the right pie chart shows a breakdown of missed objects.}
\end{figure}

\begin{figure*}
    \centering
    \includegraphics[width=400pt]{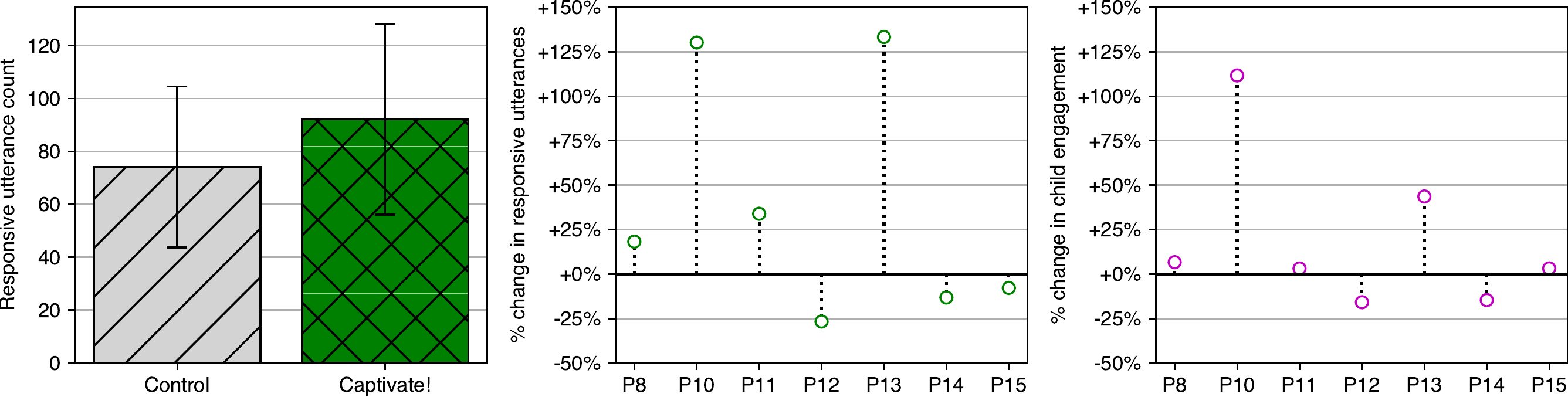}
    \caption{(Left) Total number of parent responsive utterances
        and (middle) \% differences per participant. (Right) The \% difference
        in the duration of child engagement per participant.}
    \label{fig:results_response_counts}
    \Description{A grid of three bar graphs. The left bar graph consists of the control and experiment group on the x-axis and the responsive utterance count on the y-axis. The control group averaged 79.2 and the experiment group averaged 32.6. The middle and right bar graph consists of the control and experiment group on the x-axis and the percentage change in responsive utterances and percentage change in child engagement on each y-axis.}
\end{figure*}

\begin{figure}
    \centering
        \centering
        \includegraphics[width=1.0\columnwidth]{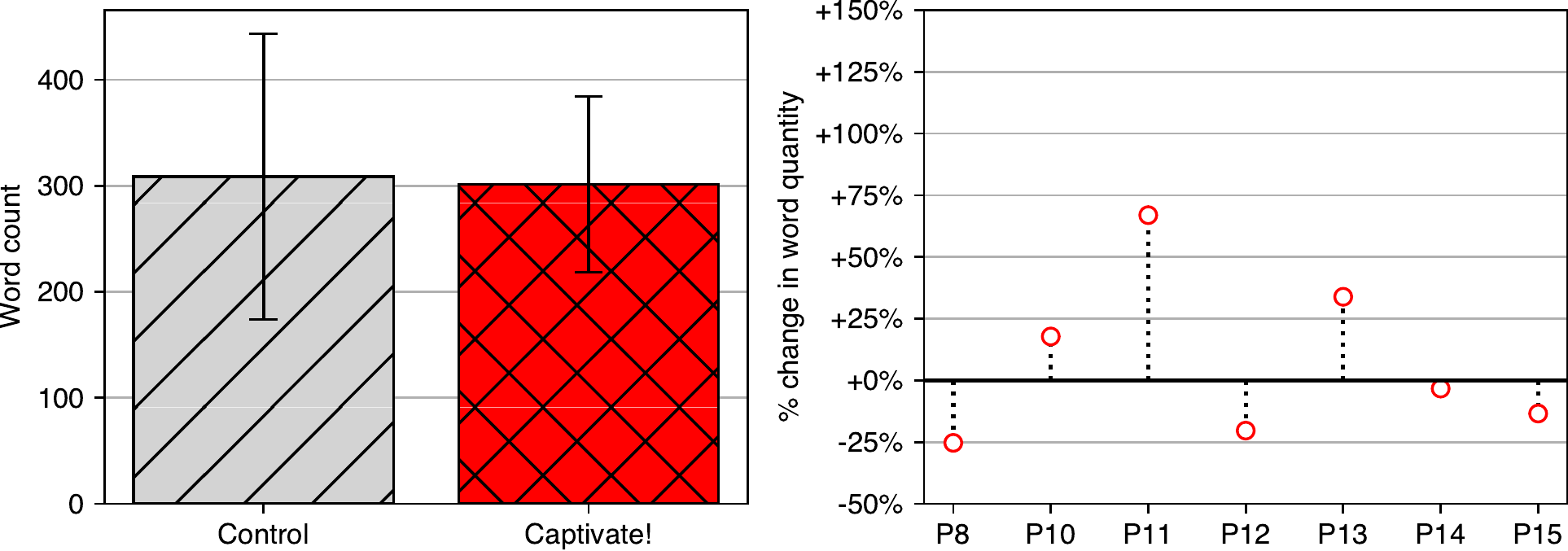}
        \caption{Number of child-directed words spoken.}
        \label{fig:results_word_quantity}
        \Description{A grid of two bar graphs. The left bar graph consists of the control and experiment group on the x-axis and the word count on the y-axis. The control group averaged 308 and the experiment group averaged 301. The right bar graph consists of the list of participants on the x-axis and the percentage change in word quantity on the y-axis.}
\end{figure}
        
\begin{figure}
    \centering
    \includegraphics[width=1.0\columnwidth]{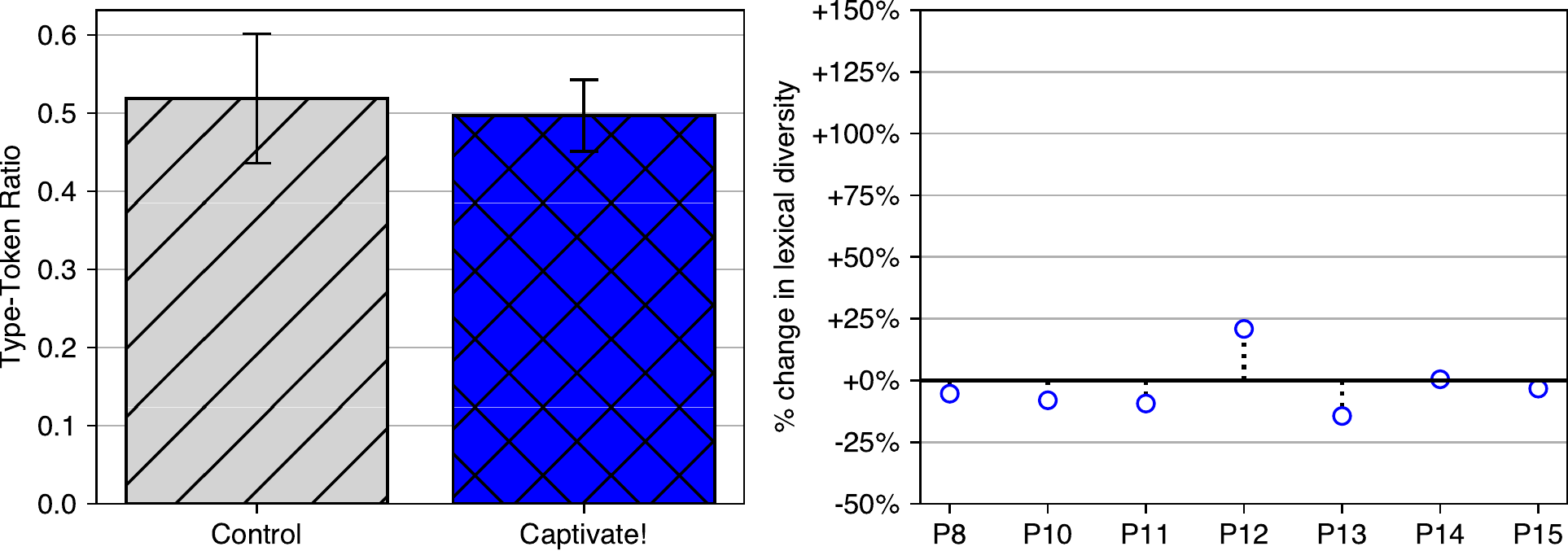}
    \caption{Lexical diversity as measured by the type--token ratio.}
    \label{fig:total_words}
    \Description{A grid of two bar graphs. The left bar graph consists of the control and experiment group on the x-axis and the type-token ratio on the y-axis. The control group averaged 0.52 and the experiment group averaged 0.50. The right bar graph consists of the list of participants on the x-axis and the percentage change in word diversity on the y-axis.}
\end{figure}

\subsection{Effects on responsiveness}
As shown in Figure \ref{fig:results_response_counts}, on average
the number of verbal responses to the child's focus of attention
increased by 38.3\% per parent when using \system{} 
compared to the control.
When viewed on a case-by-case basis---which we believe is
more appropriate due to the large variation in parent--child
interaction styles---we observed that two parents (P10, P13) 
showed extremely high increases, one (P11) showed a moderate
increase, and one (P12) showed a moderate decrease, while the
effects for the others were less significant.
Through reviewing the footage from the sessions, we could
attribute the improvement in responsiveness to higher
parent engagement towards the child, which in turn
led to higher child engagement throughout the play.
When using the cards, parents were mostly only able to focus
on the cards that they selected, which they tried
to force into the play situation.
For example, P8 went through the cards one by one.
\textit{"C8, the fish swam. look at this [card], \dots
Shall we take a look at this again?"}---although the
child was no longer interested in the fish.

\begin{figure*}
    \centering
    \includegraphics[width=\textwidth]{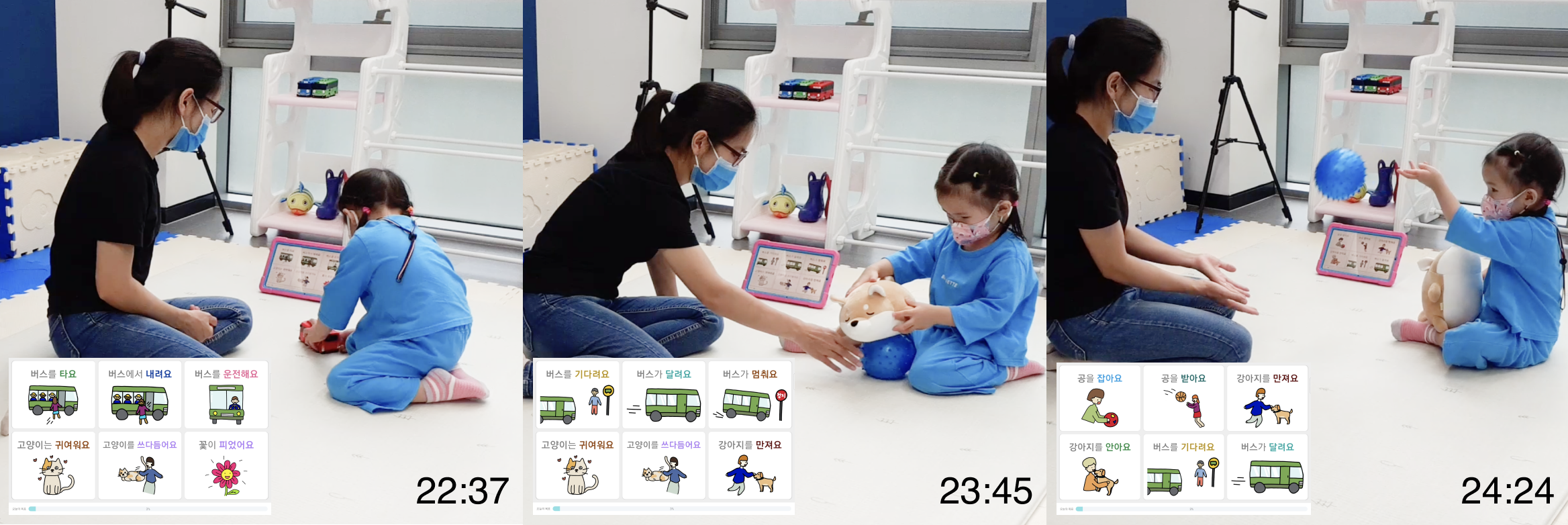}
    \caption{Example of a dynamic play situation where the 
    child's interest shifts from a bus, dog, to ball
    within two minutes. In the middle, a dog-related card 
    emerges by replacing the flower card. 
    We can observe that the final
    screen reflects all three interests.}
    \label{fig:results_samples}
    \Description{A grid of three snapshots in an experiment. There is an application screen for the corresponding scene in the lower left of each photo, and a timestamp in the lower right.}
\end{figure*}

On the other hand, the app's automated guidance
presented parents with several situation-relevant
choices, enabling them to use the target words without 
having to manually navigate a list of information. 
For example, when one child
started playing with a bus, the app increased the number of
bus-related phrases to three (Figure \ref{fig:results_samples}). When
she shifted her focus to a dog, then a ball---all within
two minutes---the app was responsive to the change, and
enabled the parent to respond with target words while also
staying engaged in the interaction. Indeed, the right of
Figure \ref{fig:results_response_counts} shows a visible
correlation between parent responsiveness and child engagement.
When using the app, the
parent was able to follow a child moving around the
room and engage in play; on contrary, when using the
cards, the parents mostly sat statically near where
the cards were placed on the ground.
This suggests that contextual language
guidance may benefit parents regardless
of their language proficiency.
As an example, 
P10---a native Korean speaker---was 
among those whose interaction
improved significantly
with \system{}.

\subsection{Effects on quantity and diversity of words}
Meanwhile, the effects on the quantity and diversity of parent 
language were less significant (+8\% in quantity, -3\% in diversity). 
The result seemed counterintuitive
considering that the number of verbal responses to the
child's focus of attention had increased significantly; however,
the footage revealed that much of the dialogue by the
parent did not follow the child's focus.
For example, P12 once saw that the child was interested 
in the fish toy and looked for a fish-related card, 
but it took more than 10 seconds to find it.
In the meantime, the child's interest had already left. 
In effect, the parents put in similar efforts to
use active and diverse language when using the cards
and \system{}---but their words were significantly more
relevant and engaging when using the app, as shown
by the higher response counts and child engagement durations.
We also note that
the parents were clearly instructed beforehand to
focus on playing while using the cards only as a reference.
However, we found that most parents wanted to teach
their children new words and actively used the cards
(for the majority of each session)
to do so;
in our exit interview, all parents reflected
that they felt both the cards and app
to be useful.

\subsection{Contextual recommendation accuracy}
\label{subsec:recommendation_accuracy}
Finally, we also measured the accuracy and sensitivity
of \system{} by marking whether the app displayed
at least one relevant card to the dominant target of
focus in every minute. As shown in Figure \ref{fig:results_system}, 
we found that a relevant
card was correctly displayed in 84.4\% of cases.
Much of the error in the missed cases was due to
occlusions; the spoon showed the highest error as it
is small and easily hidden by the child or parent
holding it in their hands.
Despite the high accuracy within one minute, 
we found that several parents expected quicker
responses when their child rapidly changed toys,
and sometimes believed it was their own fault. 
For example, P11 thought that 
\textit{``If my pronunciation is not right, 
the app doesn't work well''}. 
Such cases revealed an important
consideration when using AI technology for guidance. We discuss the implications in depth in the
next section.

%% file: sections/9_discussion.tex
\section{Discussion}
We discuss aspects of our work that may be
relevant to designers of contextual guidance systems
and technology for linguistically diverse families.

\subsection{Research challenges with linguistically diverse families}
One limitation of the evaluation was the small sample size.
Although we spent considerable effort in recruiting participants,
traditional recruitment channels were unable to
reach immigrant parents. Even with the help of
three public centers, only a small number of
parents were able to be contacted. We note that studies
with similarly minority populations tend to have smaller-sample
evaluations \cite{hwang2014talkbetter, song2016talklime, light2015designing}.
The small sample size combined with the relatively large
variation in child ages (19--47 months), parent Korean
proficiency, interaction styles, and child mood led to
large variance in the evaluation results. We
would therefore emphasize that the evaluation results
serve mainly as an early demonstration of the potential of 
contextual language guidance to aid parents.

Another challenge throughout the design process 
lied in sharing ideas across 
language barriers.
As we included immigrants from all countries in our
study, it was not feasible to prepare translation 
resources for all participants.
Consequently, in several of our interviews, we could
not directly receive detailed feedback from the parent because
they were not fluent in Korean while we did not know
the languages that they were more comfortable in.
We undertook several measures for these cases, such as
allowing the parents to invite someone who could translate; 
we also told participants
that they were free to speak in a comfortable language at any
point, as we could later translate the recording.
These measures were helped communication in many cases: 
for example,
P1 used English in her interview; P8 and P11
conducted the interview with the help of a family member.

However, we still encountered challenging interviews.
In these cases we inevitably placed more weight on
our own observations relative to the user feedback
received. Insights from experts with extensive experience 
interacting with such families were also valuable in filling
the gaps in our understanding. 
This was, of course, a compromise---ideally design research
should aim to hear detailed feedback from the users themselves.
There are potential techniques that might have
been effective in these cases, such as requesting written 
feedback in the participant's comfortable language to later
translate. We emphasize that researchers designing for 
linguistically diverse participants should think extensively
about potential challenges in communication and ways to
mitigate them.
Moreover, with designs involving children, especially infants or
toddlers, additional considerations must be made---such
as having an interview partner play 
with the child to minimize the parent's distraction during
communication.

\subsection{Sociotechnical implications of guidance technology}
Technology that prescribes \textit{guidance} must
be especially mindful of the subtle messages
that it may convey to its users.
For instance, we mentioned in Section
\ref{subsec:recommendation_accuracy} that some parents
believed it was the fault of their Korean pronunciation
when the app did not quickly respond to contextual changes.
In reality, limitations in the underlying technology
induced unavoidable errors and delays in contextual 
sensing---even for native speakers.
However, immigrant parents were more inclined to blame
their own pronunciation rather than the technology.
This may have in part reflected their experiences with 
trained AI technology such as speech recognition, 
which has been found to perform poorly for
cultural minorities \cite{koenecke2020racial}.
Indeed, we found speech recognition accuracy to
vary widely across parents.
However, as technology designers, it is imperative that 
we are aware of the sociotechnical biases that users 
perceive, and take care not to exacerbate them.
For the adoption of AI-based contextual systems such as \system{}, 
interfaces should be explored
so that users do not feel ``judged'' by
technology even in the face of errors.

Lastly, we also propose a need for exploring 
\textit{multilingual} guidance interfaces.
Although concerns about subtractive bilingualism
pressure many parents to speak to their child only in 
a less-comfortable language, this practice is being
increasingly discouraged by language development
experts \cite{baralt2020hablame}, including the SLPs we 
interviewed.
We fully support the movement towards multicultural
and multilingual homes that can bring unique growing
experiences to children.
However, guidance technologies such as \system{} may
amplify existing biases when only supporting the dominant
language.
We invite the HCI
community to explore interfaces that can support 
linguistically diverse homes, such as, potentially, bilingual
interfaces or linguistic scaffolds incorporated into the user 
experience.
Such interfaces also have the potential to include other 
caregivers such as native-born spouses in collaboratively 
fostering a rich multicultural home environment.

\subsection{Adopting contextual guidance systems}
While contextual understanding can be a powerful tool in
for guidance technology, there remain technical 
and design challenges that must be addressed for wider adoption.   
A limitation of our implementation was that the system
only recognizes, and provides guidance for, the 11 predefined objects. 
Adding support for more would require training the object
detection model on a larger dataset, as well as creating
corresponding phrase cards for each addition. While the latter
can be done manually as in our study, it is important
to also explore more scalable mechanisms for
extending guidance to new settings. One possible direction
may be in utilizing semantic networks or distributional word embeddings
to find target words for arbitrary objects \cite{grover2019semantics}.

Another consideration is in the context-sensing hardware itself.
We had chosen smartphone cameras and microphones as they
are commodity devices. 
Furthermore, they emulate key mechanisms through 
which people sense attention---by
visually observing and listening to the opponent.
We observed that gaze tends to fluctuate among 
many interests while speech is a stronger signal of
dedicated attention. Therefore, by combining these modalities,
a more comprehensive understanding of joint attention can be reached.
Three cameras were used in our 
evaluation---primarily to stabilize the accuracy of AI 
models---which would be infeasible in most home settings.
However, we expect that progress in AI such as in
more robust detection and gaze following models, as well
as face-tracking cameras, will enable
\system{} to be run entirely on a single device.
It is also possible to consider sensing hardware such
as gyroscopes attached to the toys that can detect when
a child or parent is holding it. One challenge in this case
would be in distinguishing between 
passive holding versus actively attending to the object.

With the rapid advance of AI technology,
the future holds a wealth of opportunities for contextual guidance. 
A key challenge lies in understanding
nuanced contextual information as a human can.
For example,
we once observed a child designating a boat as an ambulance because
it was red-colored; however, in this case \system{} can only 
designate the context as boat-related. 
As AI models become able to understand
diverse interactions, we believe that future interactive systems 
will be able to provide 
diverse guidance---not limited to language---in 
creative play situations such as building, drawing, or role-playing.

%% file: sections/10_conclusion.tex
\section{Conclusion}
We presented \system, the first contextual language guidance system 
that helps parents provide effective language input to their children
during interaction. 
Starting from an exploration
of language acquisition literature and an interview
with experts of immigrant language guidance, 
we established design implications for a context-driven
language guidance technology.
We addressed these implications through a user-centered 
design process with immigrant parent--child dyads and SLPs, 
which led to building joint-attention-aware guidance
with tablet-based phrase cards. 
We evaluated \system{} on parents and children aged 1--3 
to find that contextual guidance
helps parents stay responsive to their child's actions
while engaging in active and diverse dialogue.
Finally, we discussed research challenges,
sociotechnical implications, and adoption considerations
for such technology.
We conclude with an invitation for the HCI community
to explore new directions in contextual guidance 
technology for families.

%% file: sections/11_appendix.tex
\appendix

\section{Coding procedure}
\label{sec:appendix_coding}
Following Haebig et al. \cite{haebig2013contribution}, we 
extracted a 10-minute segment from each clip for coding. 
Accounting for the time needed for the participants to
adjust to the new tool, we sampled each segment 5 minutes after
the start of each session. Next, we excluded intervals where
the parent or child was distracted due to unpredictable
circumstances, such as when the parent was on a phone call, 
when the child needed a diaper change, or when the child threw
a prolonged tantrum. Finally, we extracted the metrics
described in Section \ref{sec:evaluation_procedure}.
To measure the responsiveness of the
parent throughout the interaction, we coded verbal
responses to the child's focus of attention or communication acts. 
Following \cite{haebig2013contribution}, our coding procedure followed
a two-step process. First, segments of child engagement (e.g.,
a child looking at, pointing towards, or manipulating an object) 
or communication acts (e.g., dialogue or vocalization) were labeled.
Next, within 3 seconds of the segments where the child was engaged
or had communicated, all relevant parent utterances were marked.
For example, if a child was actively playing with a teddy bear
and the parent commented, ``Hug the bear!'', the utterance was
marked as responsive. On the other hand, if the parent tried in vain
to redirect the child's attention to a different toy, such as a bus,
the utterance was not considered responsive. 
In contrast to \cite{haebig2013contribution},
we did not  responsive acts into subcategories but instead
treated them as a single class of ``responsive utterances''.
After reviewing the coding instructions of \cite{haebig2013contribution},
the first and second authors independently coded a 10-minute
segment and observed substantial agreement in interrater reliability
(Cohen's $\kappa= 0.725$). Judging the coding instructions to be
robust, subsequent clips were split and coded independently.

\section{Additional figures}
\begin{figure}[h]
    \centering
    \includegraphics[width=0.85\columnwidth]{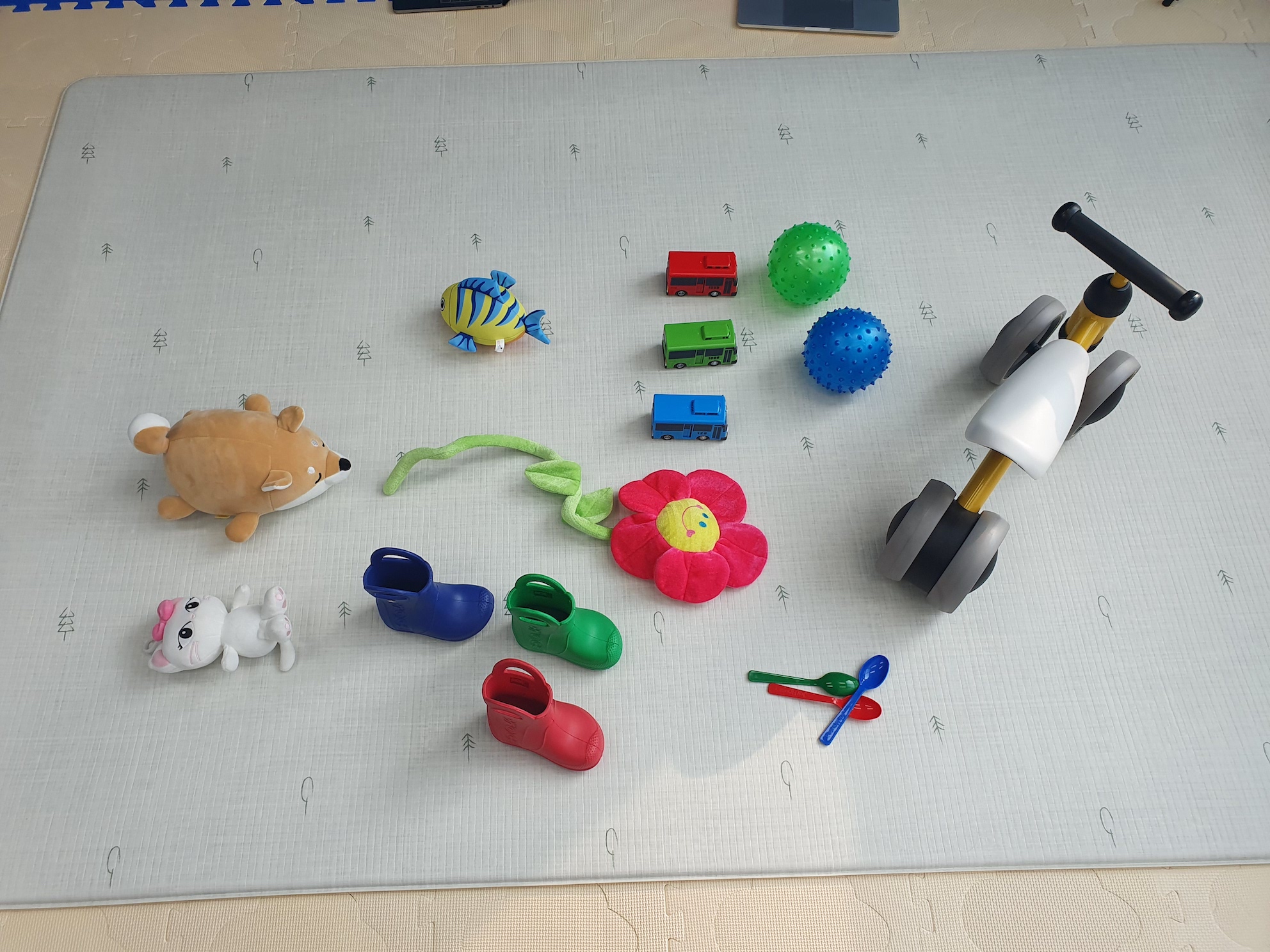} 
    \caption{Toys provided during the evaluation.}
    \label{fig:toys}
    \Description{A picture of nine kinds of toys on the floor.}
\end{figure}

\begin{figure}[h]
    \centering
    \includegraphics[width=0.85\columnwidth,height=150pt,angle=0]{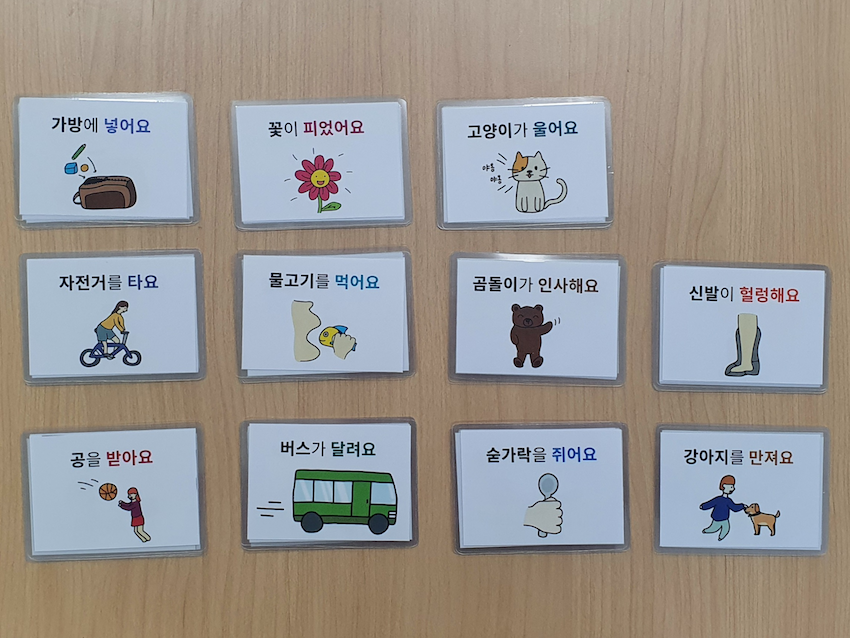} 
    \caption{Paper phrase cards.}
    \label{fig:cards}
    \Description{A picture of eleven kinds of cards on the desk.}
\end{figure}

We include additional figures about our experimental 
setup (Figures \ref{fig:toys}, \ref{fig:cards}).

\begin{figure}[h]
    \centering
    \includegraphics[width=0.85\columnwidth]{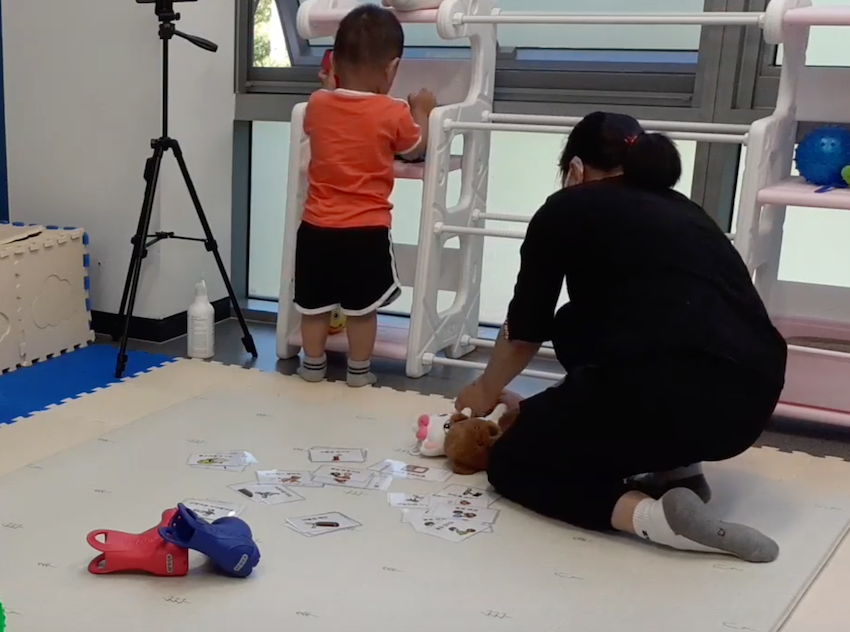} 
    \caption{A parent using paper cards in the evaluation.}
    \label{fig:card_use}
    \Description{The child and mother are turned around, cards and toys are scattered on the floor.}
\end{figure}

\section{Guiding interview questions}

\subsection{Guiding topics for the formative expert interview}
    \begin{enumerate}
	\item About the Multicultural Family Support Center and SLPs 
    \begin{itemize}
        \item Experience at Multicultural Family Support Centers
        \item Overview of language development support programs at the center
    \end{itemize}
    
    \item Differences in a language support for children in multicultural families vs. native Korean families
    \begin{itemize}
        \item Goals of the language development program
        \item Procedure of language development program
        \item Methods for evaluating the effectiveness of language development sessions
        \item Number, age, and language level of children who are participating in the language programs 
        \item Nationality and language level of the parents whose children are participating in the language programs
    \end{itemize}
    
    \item Linguistic stimulation in multicultural families
    \begin{itemize}
        \item The role and importance of language guidance at home in the development of a child's language.
        \item Difficulties commonly encountered by parents of multicultural families during language guidance at home.
        \item Effective parental language and interaction habits.
        \item The use of bilingualism in multicultural families.
    \end{itemize}
    \end{enumerate}
    
\subsection{Questions asked at post-experiment interview} 
    \begin{enumerate}
	\item Context-awareness
    \begin{itemize}
        \item Did you feel that the app responds to match the play situation?
        \item Did you think that the response speed was appropriate?
        \item Do you think that the automated interface is convenient?
    \end{itemize}
    
    \item Card vs. App
    \begin{itemize}
        \item Did you feel a difference in using the two?
        \item Is there a more comfortable tool between the two? Why?
    \end{itemize}
    
    \item Design elements of the app
    \begin{itemize}
        \item Was the read-aloud function useful?
        \item Did the picture help you understand the sentence?
        \item Was today's goal useful?
    \end{itemize}
    
    \item Usefulness of phrases
    \begin{itemize}
        \item Are phrases applicable to play situations? 
        \item If it was hard to use, why?
    \end{itemize}
    
    \item{Language interaction at home}
    \begin{itemize}
        \item Do you usually play with your child?
        \item Do you usually use a language card?
        \item What is the difference in interaction compared to the usual play situation?
    \end{itemize}
    
    \item{Parents' exposure to the target language}
    \begin{itemize}
        \item What was your previous country?
        \item How long have you lived in Korea?
    \end{itemize}
    
    \item{Children's experience in a language guidance program}
    \begin{itemize}
        \item Have you ever participated in a language guidance program at the Multicultural Family Support Center?
        \item If there is, how low was that?
    \end{itemize}
\end{enumerate}

\section{Parent informatics}
Automated tracking of attention can enable useful features 
beyond real-time guidance.
We implemented per-session statistics with (1) a 
breakdown of the toys attended throughout the play, and 
(2) how frequently the parent had used each target word 
in their dialogue. In the long term, these statistics can
reveal insights into the development of a child's interests or a
parent's language habits. However, as the
\system{} prototype was used by each parent only in a single
session, the usefulness of long-term statistics was not explored
in this study. That said, with the rise of personal 
informatics (PI) technology, we
believe that automated tracking of long term parent--child
interaction is a promising direction for exploration.